\documentclass[11pt]{article}

\usepackage{amsmath,amssymb}
\usepackage{cite}
\usepackage[dvips]{color,graphicx}
\begin{document}
\title{\Large\bf Kaon-condensed hypernuclei as highly dense self-bound objects}
\author{Takumi Muto\thanks{Email address: takumi.muto@it-chiba.ac.jp}\\
Department of Physics, Chiba Institute of Technology \\
        2-1-1 Shibazono, Narashino, Chiba 275-0023, Japan}

\date{\today}
\maketitle

\vspace{1.0cm}
      
\begin{abstract}
The structure of $K^-$-condensed hypernuclei, which may be produced in the laboratory in strangeness-conserving processes, is investigated using an effective chiral Lagrangian for the kaon-baryon interaction, combined with a nonrelativistic baryon-baryon interaction model. It is shown that a large number of negative strangeness is needed for the formation of highly dense and deeply bound state with kaon  condensates and that part of the strangeness should be carried by hyperons mixed in the nucleus. 
The properties of kaon-condensed hypernuclei such as the ground state energy and particle composition are discussed. Such a self-bound object has a long lifetime and may decay only through weak interaction processes. Comparison with other possible nuclear states is also made, such as kaon-condensed nuclei without mixing of hyperons and noncondensed multistrange hypernuclei. Implications of kaon-condensed hypernuclei for experiments are mentioned.
\end{abstract}

\begin{description} 
{\footnotesize\item PACS: 05.30.Jp, 13.75.Jz, 26.60.+c, 95.35.+d 
\item Keywords:  kaon-baryon interaction, kaon condensation, hyperons, self-bound objects}
\end{description}

\newpage
\section{Introduction}
\label{sec:intro}

\ \ Strangeness degree of freedom manifests itself in various phases of hadronic matter at high density 
and/or high temperature. It provides us with a clue to understand high-density QCD from both hadronic and quark-gluon phases\cite{ke07}. 
Kaon dynamics in a hadronic medium is an important subject associated with role of strangeness in highly dense matter\cite{f06}. In neutron stars, kaon condensation may exist as a macroscopic realization of strangeness (Bose-Einstein condensation)\cite{kn86,mtt93,lbm95,fmmt96,pbp97}. In kaon-condensed matter,  the net strangeness is produced through weak interaction processes, which play a decisive role on the ground state properties of the kaon-condensed phase in chemical equilibrium for weak processes such as $n\rightleftharpoons p+K^-$, $e^- \rightleftharpoons K^-+\nu_e$\cite{mt92}. 
The existence of kaon condensation leads to softening of the equation of state (EOS) and has a sizable effect on the internal structure of neutron stars\cite{mtt93,lbm95,fmmt96,pbp97,tpl94}. Kaon condensation is also important for thermal evolution of neutron stars since the neutrino emission processes are largely enhanced in the presence of kaon condensates\cite{bk88,t88,fmtt94}.
In recent studies, effects of the phase-equilibrium condition associated with the first-order phase transition 
on the nonuniform structure of kaon-condensed phase in neutron stars have also been pointed out\cite{g01,hps93,cgs00,nr01,vyt03,mtv06}.

Meanwhile, several works on coexistence or competition of kaon condensation with hyperons in neutron stars have been elucidated\cite{m93,kvk95,ekp95,sm96,pbg00,bb01,m01,kv03,mpp05,rhhk07,m02,m07}. The onset density and the EOS for the kaon-condensed phase in hyperonic matter have been discussed with the relativistic mean-field theory\cite{ekp95,sm96,pbg00,bb01}, the quark-meson coupling models\cite{mpp05,rhhk07}. Recently the in-medium kaon dynamics and mechanisms of kaon condensation stemming from the $s$ and $p$-wave kaon-baryon interactions in hyperonic matter have been investigated\cite{m01,m02,kv03}. 

Kaon dynamics in high density hadronic matter has also been studied through experimental achievement. In particular, deeply bound kaonic nuclear states have been proposed based on the strongly attractive antikaon-nucleon interaction\cite{ay02,yda04}. Subsequently, searching for such kaonic nuclei and $\bar K$ nuclear clusters as a cold and highly dense system has been elaborated both theoretically and experimentally\cite{k05,i03,a05,mfg05,gfgm07,y05,cvgg06,zpln06,amr06,kh07,is07,sgm07,dw07}.
These studies of the kaonic nucleus may in turn provide us with information on kaon condensation in neutron stars.

Recently, we have discussed coexistence of kaon condensation with hyperonic matter, where hyperons 
($\Lambda$, $\Sigma^-$) are mixed in addition to the neutrons, protons and leptons in the ground state of neutron-star matter\cite{m02,m07}. We pointed out that both the hyperon-mixing effect and the $s$-and $p$-wave (anti) kaon-baryon attractions soften the EOS of the kaon-condensed phase in hyperonic matter considerably, while the stiffness of the EOS is recovered at high densities due to the repulsive interaction between baryons. As a result, the energy of the system has a local minimum, which suggests the existence of the self-bound object with kaon condensates as a density isomer on any scale from an atomic nucleus to a neutron star. 
Based on the EOS, we have discussed the structure of the kaon-condensed neutron star as a self-bound star and its implications for observations of mass-radius relations and a possibility for baryonic dark matter\cite{m02,m07}. 

As for the self-bound states which may be created and exist in the early universe, compact stars, and heavy-ion collisions, a possibility of strange matter has been proposed as the true ground state or metastable state of matter\cite{i70,b71,ck79,w84,fj84}. Several hadronic phases such as abnormal states\cite{lw74}, those with pion condensation\cite{h75,m76} or $K^0$ condensation\cite{lnt90}, and hyperonic matter\cite{sgs92,shsg02, wssz99,sl99} have also been discussed as candidates for absolutely stable states. These absolutely stable states may form a bound state with strong interactions, leading to a self-bound object on an arbitrary scale, essentially without gravitation. 

In this paper, we extend the results of the self-bound object coming from the coexistence of kaon condensates with hyperonic matter to a strangeness-conserving system, and we consider a kaon-condensed hypernucleus  (abbreviated as KCYN) that is expected to be created in laboratories as a new candidate for the self-bound object. Role of hyperon-mixing on the ground state properties of the KCYN is clarified.\footnote{Early result of this work has been partly reported in Ref.~\cite{m07-2}.}

The paper is organized as follows. In Sec. 2, we take an overview of the KCYN as a strangeness-conserving system in a liquid-drop picture. Section 3 gives the expression of the volume part of the energy for the KCYN. The framework based on chiral symmetry for the kaon-baryon dynamics coupled to the baryon-baryon interaction model is presented. Numerical results are shown in Secs.~4 and 5. In Sec.~4, we discuss the structure of the KCYN such as the ground state energy, particle composition for a typical set of baryon number, electric charge, and negative strangeness. 
Hyperon-mixing effects are also addressed. 
In Sec.~5, the dependence of properties of the KCYN on the strangeness number is discussed. Comparison with the other possible nuclear states is also made such as the kaon-condensed nucleus without mixing of hyperons and noncondensed multistrange hypernucleus. Implications of the KCYN for experiments are mentioned in Sec.~6. 
Section 7 is devoted to a summary and concluding remarks. 

\section{Description of the kaon-condensed hypernucleus}
\label{sec:2}

\subsection{Liquid-drop picture}
\label{subsec:2-1}

\ \ An initial system is taken to be a target nucleus with mass number $A$ and atomic number $Z$ and the antikaons ($K^-$), the number of which is $|S|$ (the total negative strangeness). The $K^-$s' are embedded in the nucleus, and the multi-antikaonic bound state (KCYN) with the baryon number $A$, the electric charge $Z-|S|$, and strangeness $S$ is supposed to be ultimately formed. In the ground state of the KCYN, part of the strangeness $S$ should be carried by hyperons, which are mixed in the nucleus through the strong processes, $K^- N\rightarrow \pi Y$, $K^- Y\rightarrow \pi Y'$. In this paper, we take into account the $\Lambda$, $\Xi^-$, and $\Sigma^-$ for  hyperons as the constituents of the KCYN.

A spherical liquid-drop picture for the KCYN with radius $R_A$ is adopted, and uniform distributions of the baryons are assumed inside the nucleus. 
 The uniform baryon number density inside the KCYN is denoted as $\rho_{\rm B}$.  Then $R_A$ and $\rho_{\rm B}$ are related by $\displaystyle \frac{4}{3} \pi R_A^3\rho_{\rm B}=A$. Following the assumption of the uniform distribution of baryons,  
the classical $K^-$ field in the nonlinear representation within the framework of chiral symmetry is taken to be uniform:
\begin{equation}
\langle K^-\rangle =\langle K^-|\hat K^-|K^-\rangle=\frac{f}{\sqrt{2}}\theta\exp(-i\mu_{K^-} t) \ , 
\label{eq:kfield}
\end{equation}
where $\theta$ is the chiral angle, $\mu_{K^-}$ the chemical potential of the condensed $K^-$, and $f$ ($\simeq$ 93 MeV) is the meson decay constant. The type of the classical $K^-$ field, [(\ref{eq:kfield})] is the same as that for the $s$-wave kaon condensation in infinite matter, where the spatial momentum of kaon is zero.

The electric charge, strangeness number, and baryon number conservations are imposed on the KCYN:
\begin{subequations}\label{eq:conv}
\begin{eqnarray}
& &\rho_p -\rho_{\Xi^-}-\rho_{\Sigma^-}-\rho_{K^-}=\rho_{\rm B}\cdot (Z-|S|)/A \ , \label{eq:conv1} \\
& &\rho_{K^-}+\rho_\Lambda +2\rho_{\Xi^-}+\rho_{\Sigma^-}
=\rho_{\rm B} \cdot |S|/A \ , \label{eq:conv2} \\
& &\rho_p+\rho_\Lambda+\rho_{\Xi^-}+\rho_{\Sigma^-}+\rho_n=\rho_{\rm B} \ , \label{eq:conv3}
\end{eqnarray}
\end{subequations}
in terms of the number density $\rho_a$ of the particle $a$ ($a=K^-, p, \Lambda, \Xi^-, \Sigma^-, n$).
We construct the effective Hamiltonian density by introducing the electric charge chemical potential $\mu_Q$, the strangeness chemical potential $\mu_s$,  and the baryon number chemical potential $\nu$,  as the Lagrange multipliers corresponding to these  conditions. 
The resulting effective energy density is written in the form 
\begin{eqnarray}
{\cal E}_{\rm eff}={\cal E}_{\rm vol}(\theta, \lbrace\rho_a\rbrace)+\mu_Q (\rho_p-\rho_{\Xi^-}-\rho_{\Sigma^-}-\rho_{K^-})&+&\mu_s(\rho_{K^-}+\rho_\Lambda +2\rho_{\Xi^-}+\rho_{\Sigma^-}) \cr
&+&\nu (\rho_p+ \rho_\Lambda+\rho_{\Xi^-}+\rho_{\Sigma^-}+\rho_n) \ , 
\label{eq:eff}
\end{eqnarray}
where ${\cal E}_{\rm vol}(\theta, \lbrace\rho_a\rbrace)$ ($a=K^-, p, \Lambda, \Xi^-, n, \Sigma^-$) is the volume part of the energy density of the KCYN (see Sec.~\ref{sec:3}).
The conditions (\ref{eq:conv1}),(\ref{eq:conv2}), and (\ref{eq:conv3}) are consistently expressed as $\partial{\cal E}_{\rm eff}/\partial\mu_Q =\rho_{\rm B}\cdot (Z-|S|)/A$,  $\partial{\cal E}_{\rm eff}/\partial\mu_s =\rho_{\rm B}\cdot |S|/A$, and $\partial{\cal E}_{\rm eff}/\partial\nu=\rho_{\rm B}$, respectively. 
From the extremum conditions for ${\cal E}_{\rm eff}$ with respect to variation of $\rho_i$, one obtains 
\begin{subequations}\label{eq:cep}
\begin{eqnarray}
\mu_{K^-}&=&\mu_Q-\mu_s \ , \label{eq:cepk} \\
\mu_p &=&-(\mu_Q+\nu) \ , \label{eq:cepp} \\
\mu_\Lambda &=&-(\mu_s+\nu) \ , \label{eq:cepl} \\
\mu_{\Xi^-} &=&\mu_Q-2\mu_s-\nu \ , \label{eq:cepxi} \\
\mu_n &=&-\nu \ , \label{eq:cepn} \\
\mu_{\Sigma^-} &=&\mu_Q-\mu_s-\nu \ , \label{eq:ceps}
\end{eqnarray}
\end{subequations}
where $\mu_a$ (=$\partial{\cal E}_{\rm vol}(\theta, \lbrace\rho_a\rbrace)/\partial\rho_a$) is the chemical potential of the particle $a$. From Eqs.~(\ref{eq:cep}), one obtains
\begin{equation}
\mu_{K^-}=\mu_\Lambda-\mu_p=\mu_{\Xi^-}-\mu_\Lambda=\mu_{\Sigma^-}-\mu_n \ .
\label{eq:chemeq}
\end{equation}
Equation~(\ref{eq:chemeq}) ensures that the system is in chemical equilibrium for the strong processes, $K^-p\rightleftharpoons \Lambda$, $K^-\Lambda\rightleftharpoons \Xi^-$, $K^-n\rightleftharpoons \Sigma^-$.\footnote{We assume that the pions which may be produced through the processes, $K^-N\rightarrow \pi Y$, $K^-Y\rightarrow \pi Y'$ above the pion threshold, are emitted outside the nucleus, so that they don't take part in the chemical equilibrium for the relevant processes.} 
These conditions should be compared with those for kaon condensation realized in neutron stars, where the total strangeness is produced through the weak processes such as $n\rightleftharpoons pK^-$, $n\rightleftharpoons pe^-(\bar\nu_e)$, $ne^-\rightleftharpoons \Sigma^-(\nu_e)$, and $n\rightleftharpoons \Lambda(\nu_e\bar\nu_e)$, and the system is in chemical equilibrium for these weak processes\cite{m07}.

By the use of Eqs.~(\ref{eq:cep}), the effective energy density ${\cal E}_{\rm eff}$ is represented as
\begin{equation}
{\cal E}_{\rm eff}(\theta, \lbrace\mu_a\rbrace)
={\cal E}_{\rm vol}(\theta, \lbrace\rho_a\rbrace)-\sum_{a=K^-,p,\Lambda,\Xi^-,n,\Sigma^-}\mu_a\rho_a \ .
\label{eq:eff2}
\end{equation}

\subsection{Surface and Coulomb energy effects}
\label{subsec:2-2}

\ \ In a similar way to the nuclear mass formula, the total energy of the KCYN at zero temperature is written as
\begin{equation}
 E(\theta, \lbrace\rho_a\rbrace; A, Z-|S|,|S|)= A\cdot {\cal E}_{\rm vol}(\theta, \lbrace\rho_a\rbrace)/\rho_{\rm B}
 +4\pi R_A^2\sigma+\frac{3}{5}\frac{(Z-|S|)^2 e^2}{R_A}\ , 
\label{eq:energy}
\end{equation}
where the second term on the r.h.s. of Eq.~(\ref{eq:energy}) is  the surface energy with $\sigma$ being the surface tension coefficient, and the third term is the bulk Coulomb energy. The surface energy coefficient  is assumed to be similar to the value deduced from the empirical mass formula of the normal nucleus and is set to be $\sigma $ = 1 MeV/fm$^2$\cite{rs80}.

\section{Volume part of the energy}
\label{sec:3}

\subsection{Kaon-baryon interactions from chiral symmetry}
\label{subsec:chiral}

\ \ For the volume energy part ${\cal E}_{\rm vol}(\theta, \lbrace\rho_a\rbrace)/\rho_{\rm B}$ , the $s$-wave kaon-baryon interactions are incorporated from the effective chiral Lagrangian. 

\begin{eqnarray}
{\cal L}&=&\frac{1}{4}f^2 \ {\rm Tr} 
\partial^\mu\Sigma^\dagger\partial_\mu\Sigma 
+\frac{1}{2}f^2\Lambda_{\chi{\rm SB}}({\rm Tr}M(\Sigma-1)+{\rm h.c.}) 
\cr
&+&{\rm Tr}\overline{\Psi}(i{\not\partial}-M_{\rm B})\Psi+{\rm 
Tr}\overline{\Psi}i\gamma^\mu\lbrack V_\mu, \Psi\rbrack
\cr
&+&a_1{\rm Tr}\overline{\Psi}(\xi M^\dagger\xi+{\rm h.c.})\Psi 
+ a_2{\rm Tr}\overline{\Psi}\Psi(\xi M^\dagger\xi+{\rm h.c.})+
a_3({\rm Tr}M\Sigma +{\rm h.c.}){\rm Tr}\overline{\Psi}\Psi \ , 
\label{eq:lag}
\end{eqnarray}
where the first and second terms on the r.~h.~s. of Eq.~(\ref{eq:lag}) are the kinetic and mass terms of mesons, respectively. $\Sigma$ is the nonlinear meson field defined by $\Sigma\equiv e^{2i\Pi/f}$, where $\displaystyle\Pi\equiv\sum_{a=1\sim 8}\pi_aT_a$ with $\pi_a$ being the octet meson fields and $T_a$ being the SU(3) generators. 
Since only charged kaon condensation is considered, the $\Pi$ is simply written as 
\begin{eqnarray}
\Pi=\frac{1}{\sqrt{2}}\left(
\begin{array}{ccc}
0 & 0 & K^+ \\
0 & 0 & 0 \\
K^- & 0 & 0 \\
\end{array}\right) \ .
\label{eq:meson}
\end{eqnarray}
In the second term on the r.~h.~s. of Eq.~(\ref{eq:lag}), $\Lambda_{\chi{\rm SB}}$ is the chiral symmetry breaking scale, $\sim$ 1 GeV, $M$ the mass matrix which is given by 
$M\equiv {\rm diag}(m_u, m_d, m_s)$ with the quark masses $m_i$. 
The third term in Eq.~(\ref{eq:lag}) denotes the free baryon part, where the $\Psi$ is the octet baryon field including only the $p$, $\Lambda$, $\Xi^-$, $n$, $\Sigma^-$,
and $M_{\rm B}$ the baryon mass generated as a consequence of spontaneous chiral symmetry breaking.
The fourth term in Eq.~(\ref{eq:lag}) gives the $s$-wave kaon-baryon interaction of the vector type corresponding to the Tomozawa-Weinberg term with $V_\mu$ being the mesonic vector current defined by $V_\mu\equiv 
1/2(\xi^\dagger\partial_\mu\xi+\xi\partial_\mu\xi^\dagger)$ with $\xi\equiv \Sigma^{1/2}$. 
 The last three terms in Eq.~(\ref{eq:lag}) give the 
$s$-wave meson-baryon interaction of the scalar type, which explicitly breaks chiral symmetry.

The quark masses $m_i$ are chosen to be $m_u$ = 6 MeV, $m_d$ = 12 MeV, and $m_s$ = 240 MeV. Together with these values, the parameters $a_1$ and $a_2$ are fixed to be $a_1$ = $-$0.28, $a_2$ = 0.56 so as to reproduce the empirical octet baryon mass splittings\cite{kn86}. The parameter $a_3$ is related to the kaon-nucleon ($KN$) sigma terms simulating the $s$-wave $KN$ attraction of the scalar type through the expressions, $\Sigma_{Kp}=-(a_1+a_2+2a_3)(m_u+m_s)$, $\Sigma_{Kn}=-(a_2+2a_3)(m_u+m_s)$, evaluated at the on-shell Cheng-Dashen point for the effective chiral Lagrangian (\ref{eq:lag}). 
Recent lattice calculations suggest the value of the $KN$ sigma term $\Sigma_{KN}$=(300$-$400) MeV\cite{dll96}. Throughout this paper, we take the value of $a_3=-0.9$, leading to $\Sigma_{Kn}$=305 MeV, as a standard value. 
The $K^-$ optical potential in symmetric nuclear matter, 
$V_{\rm opt}(\rho_{\rm B})$, is estimated as a scale of the $K^-$-nucleon attractive interaction. It is defined by
\begin{equation}
V_{\rm opt}(\rho_{\rm B})=\Pi_{K^-}(\omega(\rho_{\rm B}),\rho_{\rm B})
/2 \omega(\rho_{\rm B}) \ , 
\label{eq:vopt}
\end{equation}
where $\Pi_{K^-}\left(\omega(\rho_{\rm B}),\rho_{\rm B}\right)$
is the $K^-$ self-energy 
at given $\rho_B$ with $\rho_p=\rho_n=\rho_{\rm B}/2$. 
For $a_3=-0.9$, $V_{\rm opt}(\rho_0)$ is estimated to be $-$ 115 MeV at the nuclear saturation density $\rho_0$ (=0.16 fm$^{-3}$). 

 In our framework, the kaon-baryon interactions are incorporated in the lowest order in chiral perturbation except for the kaon-baryon sigma terms which are of the order $O(m_K^2)$. 
To be consistent with the on-shell $s$-wave $K$ ($\bar K$)-$N$ scattering lengths, 
we have to take into account the range terms proportional to $\omega^2$ coming from the higher-order terms in chiral expansion and a pole contribution from the 
$\Lambda$(1405) which we regard as an elementary particle. Nevertheless, these contributions to the energy density become negligible in high-density matter as far as $K^-$ condensation is concerned\cite{lbm95,fmmt96}. 
Therefore, we omit these correction terms throughout this paper and consider the simplified expression for the energy density of the kaon-condensed phase. 

Concerning the deeply bound kaonic nuclei, a deep $\bar K$ potential was proposed phenomenologically by Akaishi and Yamazaki on the assumption that the $\Lambda$(1405) is a bound state of the $K^-$ and proton\cite{ay02}. While their derivation of the $\bar K$ potential is different from ours, their potential depth at $\rho_0$ is similar to that used in this paper. 
On the other hand, Oset and Toki made a critical analysis on the predicted deeply bound kaonic nuclei on the basis of a shallow $\bar K$ potential derived from the chiral unitary model and the two-pole structure of the $\Lambda$(1405) resulting from their chiral approach\cite{ot06}. Recently, Akaishi and Yamazaki put forward a counterargument\cite{ya07}. Thus there is still controversy about $\bar K$-baryon interactions in dense matter and the possible existence of the deeply bound kaonic nuclei.

\subsection{Effective energy density}
\label{subsec:energy}

\ \ The total effective energy density ${\cal E}_{\rm eff}$ [(\ref{eq:eff2})] is separated into two parts: 
${\cal E}_{\rm eff}={\cal E}_{\rm eff}^{\rm B}+{\cal E}_{\rm eff}^{\rm M}$, where the first term on the r.h.s. is the baryon part including the kaon-baryon interaction and the second term is the meson part consisting of the free classical kaons only. The baryon part ${\cal E}_{\rm eff}^{\rm B}$ and the meson part ${\cal E}_{\rm eff}^{\rm M}$ are derived from the effective chiral Lagrangian (\ref{eq:lag}) together with the number density of the kaon condensates $\rho_{K^-}$ through the term, $-\mu_{K^-}\rho_{K^-}$ in (\ref{eq:eff2}). 
The number density of the kaon condensates $\rho_{K^-}$ is given from the time component of the mesonic part of the strangeness current associated with the Lagrangian (\ref{eq:lag}):
\begin{equation}
j_s^\mu=-i\frac{f^2}{4}{\rm Tr}(\partial^\mu\Sigma^\dagger [S, \Sigma]-{\rm h. c.})+{\rm Tr}\bar\psi\gamma^\mu[i\tilde{\cal V},\psi] \ , 
\label{eq:scurrent}
\end{equation}
where $S={\rm diag}(0,0,1)$ and 
$\displaystyle\tilde{\cal V}=-\frac{i}{2}(\xi^\dagger[S,\xi]+\xi[S,\xi^\dagger])$\cite{lnt90}. 
From Eqs.~(\ref{eq:kfield}) and (\ref{eq:scurrent}), one obtains
\begin{equation}
\rho_{K^-} = \mu_{K^-} f^2\sin^2\theta+\left(\rho_p+\frac{1}{2}\rho_n-\rho_{\Xi^-}-\frac{1}{2}\rho_{\Sigma^-}\right)(1-\cos\theta) \ .
\label{eq:rhok}
\end{equation}
It is to be noted that the baryon part ${\cal E}_{\rm eff}^{\rm B}$ includes the contribution from the kaon-baryon interaction term in $\rho_{K^-}$ [the second term on the r.h.s. of Eq.~(\ref{eq:rhok})] and that the meson part ${\cal E}_{\rm eff}^{\rm M}$ includes the contribution from the free kaon part in $\rho_{K^-}$ [the first term on the r.h.s. of Eq.~(\ref{eq:rhok})]. After the nonrelativistic reduction for the baryon part of the effective Hamiltonian by way of the Foldy-Wouthuysen-Tani transformation and with the mean-field approximation, one obtains
\begin{equation}
{\cal E}_{\rm eff}^{\rm B}=\sum_{i=p,\Lambda,\Xi^-, n,\Sigma^-} \sum_{\stackrel{|{\bf p}| \leq 
|{\bf p}_F(i)|}{s=\pm1/2}}E_{{\rm eff},s}^{(i)} ({\bf p}) \ , 
\label{eq:be}
\end{equation}
where ${\bf p}_F(i)$ is the Fermi momentum of the baryon $i$, and the 
subscript `$s$' stands for the spin states for the baryon. 
The effective single-particle energies $E_{{\rm eff},s}^{(i)} ({\bf p})$ are represented by
\begin{subequations}\label{eq:spe}
\begin{eqnarray} 
E_{{\rm eff},s}^{(p)} ({\bf p})&=& {\bf p}^2/2M_N -(\mu_{K^-}+\Sigma_{Kp})(1-\cos\theta)-\mu_p \ ,  \label{eq:spep} \\
E_{{\rm eff},s}^{(\Lambda)} ({\bf p})&=& {\bf p}^2/2M_N -\Sigma_{K\Lambda}(1-\cos\theta)
+\delta M_{\Lambda N}-\mu_\Lambda  \ , \label{eq:spel} \\
E_{{\rm eff},s}^{(\Xi^-)} ({\bf p})&=& {\bf p}^2/2M_N-\Big(-\mu_{K^-}+\Sigma_{K\Xi^-}\Big)(1-\cos\theta)
+\delta M_{\Xi^- N}-\mu_{\Xi^-} \ ,\label{eq:spexi} \\
E_{{\rm eff},s}^{(n)} ({\bf p})&=&{\bf p}^2/2M_N-\Big(\frac{1}{2}\mu_{K^-}+\Sigma_{Kn}\Big)(1-\cos\theta)-\mu_n \ , \label{eq:spen} \\
E_{{\rm eff},s}^{(\Sigma^-)} ({\bf p})&=& {\bf p}^2/2M_N-\Big(-\frac{1}{2}\mu_{K^-}+\Sigma_{K\Sigma^-}\Big)(1-\cos\theta)
+\delta M_{\Sigma^- N}-\mu_{\Sigma^-} \ , \label{eq:spes} 
\end{eqnarray}
\end{subequations}
where $M_N$ is the nucleon mass, $\delta M_{\Lambda N}$ (= 176 MeV), $\delta M_{\Xi^- N}$ (= 382 MeV), and $\delta M_{\Sigma^- N}$ (= 258 MeV) are the hyperon-nucleon mass differences. The ``kaon-hyperon sigma terms'' are defined by $\displaystyle\Sigma_{K\Lambda}\equiv -\left(\frac{5}{6}a_1+\frac{5}{6}a_2+2a_3\right)(m_u+m_s)$, $\Sigma_{K\Xi^-}\equiv -(a_1+a_2+2a_3)(m_u+m_s)$(=$\Sigma_{Kp}$),  
and $\displaystyle \Sigma_{K\Sigma^-}\equiv -(a_2+2a_3)(m_u+m_s)$ (=$\Sigma_{Kn}$). It is to be noted that each term in Eqs.~(\ref{eq:spe}) contains both the kaon-baryon attraction of the scalar type simulated by the ``sigma term'' and the kaon-baryon interaction of the vector type proportional to $\mu_{K^-}$ the coefficient of which is given by the V-spin charge of each baryon (Tomozawa-Weinberg term).  
 
The meson contribution to the effective energy density, ${\cal E}_{\rm eff}^{\rm M}$, is given by 
the substitution of the classical kaon field (\ref{eq:kfield}) 
into the meson part of the effective Hamiltonian:  
\begin{equation}
{\cal E}_{\rm eff}^{\rm M}=-\frac{1}{2}f^2\mu_{K^-}^2\sin^2\theta+f^2m_K^2(1-\cos\theta) \ , 
\label{eq:me}
\end{equation}
where $m_K\equiv [\Lambda_{\chi {\rm SB}}(m_u+m_s)]^{1/2}$, which is 
identified with the free kaon mass, and is replaced by the 
experimental value, 493.7 MeV. 

\subsection{Potential energy contributions of baryons}
\label{subsec:pot}

\ \ We take into account the baryon potential effects on the energy of the kaon-condensed state as follows. A potential energy density ${\cal E}_{\rm pot}(\lbrace\rho_i\rbrace)$ ($i=p, \Lambda, \Xi^-, n, \Sigma^-$) is introduced as a local effective baryon-baryon interaction, 
which is assumed to be given by functions of the number densities of the relevant baryons. 
 Then the baryon potential $V_i$ ($i=p, \Lambda$, $\Xi^-$, $n$, $\Sigma^-$) is defined as 
\begin{equation}
V_i=\partial{\cal E}_{\rm pot}(\lbrace\rho_i\rbrace)/\partial\rho_i \ ,
\label{eq:vi}
\end{equation}
and it is added to each effective single particle energy, $E_{{\rm eff},s}^{(i)}({\bf p})\rightarrow {E'}_{{\rm eff},s}^{(i)}({\bf p})=E_{{\rm eff},s}^{(i)}({\bf p})+V_i$. 
The total effective energy density ${\cal E}^{\rm eff}$ is modified by addition of 
the potential energy density ${\cal E}_{\rm pot}(\lbrace\rho_i\rbrace)$ and subtraction of the term $\displaystyle\sum_{i=p, \Lambda,\Xi^-, n,\Sigma^-}\rho_i V_i$ so as to avoid the double counting of the baryon interaction energies in the sum over the effective single particle energies ${E'}_{{\rm eff},s}^{(i)}({\bf p})$. 
As a result, the baryon part of the effective energy density is modified as 
\begin{equation}
{\cal E}_{\rm eff}'^{\rm B}=\sum_{i=p, \Lambda, \Xi^-, n, \Sigma^-} \sum_{\stackrel{|{\bf p}| \leq 
|{\bf p}_F(i)|}{s=\pm1/2}}{E'}_{{\rm eff},s}^{(i)} ({\bf p})+{\cal E}_{\rm pot}(\lbrace\rho_i\rbrace)-\sum_{i=p, \Lambda, \Xi^-, n, \Sigma^-} \rho_i V_i \ .
\label{eq:ebb}
\end{equation}

The expression of the potential energy density ${\cal E}_{\rm pot}(\lbrace\rho_i\rbrace)$ is taken to be the following form given in Ref.~\cite{bg97}:
\begin{eqnarray}
{\cal E}_{\rm pot}(\lbrace\rho_i\rbrace)&=&\frac{1}{2}\Big\lbrack a_{\rm NN}(\rho_{\rm 
p}+\rho_{\rm n})^2+b_{\rm NN}(\rho_{\rm p}-\rho_{\rm n})^2
+c_{\rm NN}(\rho_{\rm p}+\rho_{\rm n})^{\delta+1} \Big\rbrack\cr
&+& a_{\rm \Lambda N}(\rho_{\rm p}+\rho_{\rm n}){\rho_\Lambda}+c_{\rm 
\Lambda N}\Bigg\lbrack\frac{(\rho_{\rm p}
+\rho_{\rm n})^{\gamma+1}}{\rho_{\rm p}+\rho_{\rm 
n}+{\rho_\Lambda}}{\rho_\Lambda}
+\frac{{\rho_\Lambda}^{\gamma+1}}{\rho_{\rm p}
+\rho_{\rm n}+{\rho_\Lambda}}(\rho_{\rm p}
+\rho_{\rm n})\Bigg\rbrack \cr
&+&\frac{1}{2}\Big\lbrack a_{YY}{\rho_\Lambda}^2
+c_{\rm YY}{\rho_\Lambda}^{\gamma+1}+(a_{\rm YY}+
b_{\Xi\Xi}){\rho_{\Xi^-}}^2+c_{\rm 
YY}{\rho_{\Xi^-}}^{\gamma+1}\Big\rbrack \cr 
&+&a_{\rm \Xi N}(\rho_{\rm p}+\rho_{\rm n}){\rho_{\Xi^-}}
+b_{\rm \Xi N}(\rho_{\rm n}-\rho_{\rm p}){\rho_{\Xi^-}} \cr
&+&c_{\rm \Xi N}\Bigg\lbrack\frac{(\rho_{\rm p}
+\rho_{\rm n})^{\gamma+1}}{\rho_{\rm p}
+\rho_{\rm n}+\rho_{\Xi^-}}\rho_{\Xi^-}
+\frac{{\rho_{\Xi^-}}^{\gamma+1}}{\rho_{\rm p}
+\rho_{\rm n}+{\rho_{\Xi^-}}}(\rho_{\rm p}
+\rho_{\rm n})\Bigg\rbrack \cr
&+&a_{\rm YY}{\rho_{\Xi^-}}{\rho_\Lambda}
+c_{\rm YY}\Bigg\lbrack 
\frac{\rho_\Lambda^{\gamma+1}}{\rho_{\Xi^-}+\rho_\Lambda}\rho_{\Xi^-}
+\frac{{\rho_{\Xi^-}}^{\gamma+1}}{{\rho_{\Xi^-}}+{\rho_\Lambda}}\rho_\Lambda\Bigg\rbrack 
\cr
&+&a_{\rm \Sigma N}(\rho_{\rm p}+\rho_{\rm 
n}){\rho_{\Sigma^-}}+b_{\rm \Sigma N}(\rho_{\rm n}-\rho_{\rm 
p}){\rho_{\Sigma^-}} \cr 
&+& c_{\rm \Sigma N}\Bigg\lbrack\frac{(\rho_{\rm p}
+\rho_{\rm n})^{\gamma+1}}{\rho_{\rm p}+\rho_{\rm 
n}+{\rho_{\Sigma^-}}}{\rho_{\Sigma^-}}
+\frac{{\rho_{\Sigma^-}}^{\gamma+1}}{\rho_{\rm p}
+\rho_{\rm n}+\rho_{\Sigma^-}}(\rho_{\rm p}
+\rho_{\rm n})\Bigg\rbrack \cr
&+&a_{\rm YY}{\rho_{\Sigma^-}}{\rho_\Lambda}+c_{\rm YY}\Bigg\lbrack 
\frac{{\rho_{\Sigma^-}}^{\gamma+1}}{{\rho_{\Sigma^-}}+{\rho_\Lambda}}{\rho_\Lambda}+\frac{{\rho_\Lambda}^{\gamma+1}}{{\rho_{\Sigma^-}}+{\rho_\Lambda}}{\rho_{\Sigma^-}}\Bigg\rbrack 
\cr
&+& a_{\rm 
YY}{\rho_{\Sigma^-}}{\rho_{\Xi^-}}+b_{\Sigma\Xi}{\rho_{\Xi^-}}{\rho_{\Sigma^-}}+c_{\rm 
YY}\Bigg\lbrack 
\frac{{\rho_{\Xi^-}}^{\gamma+1}}{{\rho_{\Xi^-}}+{\rho_{\Sigma^-}}}{\rho_{\Sigma^-}}+\frac{{\rho_{\Sigma^-}}^{\gamma+1}}{{\rho_{\Xi^-}}+{\rho_{\Sigma^-}}}{\rho_{\Xi^-}}\Bigg\rbrack 
\cr
&+&\frac{1}{2}\Big\lbrack (a_{\rm YY}
+b_{\Sigma\Sigma}){\rho_{\Sigma^-}}^2
+c_{\rm YY}{\rho_{\Sigma^-}}^{\gamma+1}\Big\rbrack \ .
\label{eq:epot}
\end{eqnarray}
The terms including the exponents $\delta$ and $\gamma$ ($\delta$, $\gamma > 1$) represent the multi-body repulsive interactions between baryons. 
The parameters for the $NN$ and $YN$ parts in 
Eq.~(\ref{eq:epot}) are refitted to be consistent with the recent empirical data on the nuclear and hypernuclear properties as follows\cite{g04,fg07,ht06}: (i) 
The parameters $a_{NN}$ and $c_{NN}$ in the $NN$ part are determined to reproduce the standard nuclear saturation density $\rho_0$=0.16 fm$^{-3}$ and the binding energy $-$16 MeV in symmetric nuclear matter. With the parameters $a_{NN}$, $c_{NN}$, and $\delta$, the incompressibility $K$ in symmetric nuclear matter is obtained. The parameter $b_{NN}$ in the isospin-dependent term in the $NN$ part is related to the potential contribution to the symmetry energy $V_{\rm sym}(\rho_{\rm B})$ as $V_{\rm sym}(\rho_{\rm B})=b_{NN}\rho_{\rm B}/2$. The value of $b_{NN}$ is fixed to reproduce the empirical value of the symmetry energy at $\rho_B=\rho_0$, $\displaystyle E_{\rm sym}=\frac{3}{5}\frac{(3\pi^2\rho_0)^{2/3}}{2M_N}(1-2^{-2/3})+V_{\rm sym}(\rho_0)\sim$ 30 MeV. (ii) For the $YN$ parts, the parameters $a_{\Lambda N}$ and $c_{\Lambda N}$ are  chosen to be the same values as those in Ref.~\cite{bg97}, where the single $\Lambda$ orbitals in ordinary hypernuclei are fitted well. The depth of the $\Lambda$ potential in nuclear matter is then given as $V_{\Lambda}(\rho_p=\rho_n=\rho_0/2)=a_{\Lambda N}\rho_0+c_{\Lambda N}\rho_0^\gamma$=$-$27 MeV\cite{mdg88,ht06}.
The depth of the $\Xi^-$ potential in nuclear matter is set to be attractive, $V_{\Xi^-}(\rho_p=\rho_n=\rho_0/2)=a_{\Xi N}\rho_0+c_{\Xi N}\rho_0^\gamma$=$-$16 MeV with reference to the experimentally deduced results from the ($K^-$, $K^+$) reactions, ($-$14)$-$($-$20) MeV\cite{f98,k00}. 

The depth of the $\Sigma^-$ potential $V_{\Sigma^-}$ in nuclear matter is taken to be repulsive and strongly isospin-dependent, which is consistent with recent theoretical calculations\cite{kf00,fk01} and the phenomenological analyses on the ($K^-$, $\pi^\pm $) reactions at BNL\cite{b99,d99}, ($\pi^-$, $K^+$) reactions at KEK\cite{n02,dr04,hh05}, and the $\Sigma^-$ atom data\cite{mfgj95}: One sets $V_{\Sigma^-}(\rho_p=\rho_n=\rho_0/2)=a_{\Sigma N}\rho_0+c_{\Sigma N}\rho_0^\gamma$=23.5 MeV and $b_{\Sigma N}\rho_0$=40.2 MeV. This choice of the parameters is compatible with the values in Ref.~\cite{d99} based on the Nijmegen model F. 
(iii) For the $YY$ part, we take the same parameters as those in Ref.~\cite{bg97}, since the experimental information on the $YY$ interactions is not enough to fix  the relevant parameters precisely. 

Taking into account the conditions (i) $\sim$ (iii), we adopt the following two parameter sets throughout this paper: (A) $\delta$=$\gamma$=5/3. In this case, one obtains $K$=306 MeV, which is larger than the standard empirical value 210$\pm$30 MeV\cite{b80}. 
(B) $\delta$=4/3 and $\gamma$=2.0. From the choice $\delta$=4/3, one obtains $K$=236 MeV which lies within the empirical value. Here we abbreviate the parameter sets (A) and (B) used in the models for the baryon-baryon interactions as H-EOS (A) and H-EOS (B), respectively. 
Numerical values of the parameters are listed in Table~\ref{tab:para}. 
It is to be noted that, in the absence of hyperons, the nucleon-nucleon interaction for H-EOS~(A) gives the stiffer EOS than that for H-EOS~(B). On the other hand,   
the many-body interactions between hyperons and nucleons  become more repulsive for H-EOS (B) ($\gamma$=2.0) than for H-EOS (A) ($\gamma$=5/3) at high densities, so that the H-EOS (B) leads to stiffer EOS than H-EOS (A) at high densities. 

\begin{table}[h]
\caption{Parameters used in the potential energy density for the KCYN. (The units: $^{\rm 
a}$MeV$\cdot$fm$^3$, \  $^{\rm b}$MeV$\cdot$fm$^{3\gamma}$, \  $^{\rm c}$MeV$\cdot$fm$^{3\delta}$)}
\label{tab:para}
\begin{center}
\begin{tabular}{c|cr|cr|cr}
\hline
H-EOS & parameter & & parameter & & parameter & \\ \hline
 & $\gamma$ & 5/3 &  ${a_{\Lambda N}}^{\rm a}$ & $-$387.0 & 
${a_{YY}}^{\rm a}$ & $-$552.6 \\
 & $\delta$ & 5/3 & ${c_{\Lambda N}}^{\rm b}$ & 738.8 & ${c_{YY}}^{\rm b}$ & 1055.4   \\
(A) & ${a_{NN}}^{\rm a}$ & $-$914.2 & ${a_{\Xi N}}^{\rm a}$ &  $-$228.6 & ${b_{\Xi\Xi}}^{\rm a}$ &  0 \\
 & ${b_{NN}}^{\rm a}$ & 212.8 & ${b_{\Xi N}}^{\rm a}$ & 0 & ${b_{\Sigma\Xi}}^{\rm a}$ &  0   \\
 & ${c_{NN}}^{\rm c}$ & 1486.4 & ${c_{\Xi N}}^{\rm b}$ & 436.5  & ${b_{\Sigma\Sigma}}^{\rm a}$ & 428.4  \\
  & & & ${a_{\Sigma N}}^{\rm a}$ & $-$70.9 & &  \\
 &  & & ${b_{\Sigma N}}^{\rm a}$ & 251.3 & &  \\
  &  & & ${c_{\Sigma N}}^{\rm b}$ & 738.8 & &  \\\hline
  & $\gamma$ & 2 &  ${a_{\Lambda N}}^{\rm a}$ & $-$342.8 & 
${a_{YY}}^{\rm a}$ & $-$486.2 \\
 & $\delta$ & 4/3 & ${c_{\Lambda N}}^{\rm b}$ & 1087.5 & ${c_{YY}}^{\rm b}$ & 1553.6   \\
(B) & ${a_{NN}}^{\rm a}$ & $-$1352.3 & ${a_{\Xi N}}^{\rm a}$ & $-$203.1 &  ${b_{\Xi\Xi}}^{\rm a}$ & 0 \\
 & ${b_{NN}}^{\rm a}$ & 212.8 & ${b_{\Xi N}}^{\rm a}$ & 0 & ${b_{\Sigma\Xi}}^{\rm a}$ &  0   \\
 & ${c_{NN}}^{\rm c}$ & 1613.9 & ${c_{\Xi N}}^{\rm b}$ & 644.4  & ${b_{\Sigma\Sigma}}^{\rm a}$ & 428.4  \\
  & & & ${a_{\Sigma N}}^{\rm a}$ & $-$27.1 & &  \\
 &  & & ${b_{\Sigma N}}^{\rm a}$ & 251.3 & &  \\
  &  & & ${c_{\Sigma N}}^{\rm b}$ & 1087.5 & &  \\\hline
\end{tabular}
\end{center}
\end{table}

\subsection{Total energy expression and constraints}
\label{subsec:3-4}

\ \ The total effective energy density after inclusion of  the baryon potential energy effects, ${\cal E}_{\rm eff}'(\theta, \lbrace\mu_a\rbrace)$, is written from Eqs.~(\ref{eq:spe}), (\ref{eq:me}) and (\ref{eq:ebb}) as 
\begin{eqnarray}
{\cal E}_{\rm eff}'(\theta, \lbrace\mu_a\rbrace)&=&
{\cal E}_{\rm eff}'^{\rm B}+{\cal E}_{\rm eff}^{\rm M}\cr
&=&\frac{3}{5}\frac{(3\pi^2)^{2/3}}{2M_N} (\rho_p^{5/3}+\rho_\Lambda^{5/3}+\rho_{\Xi^-}^{5/3}+\rho_n^{5/3}+\rho_{\Sigma^-}^{5/3}) \cr
&+&(\rho_\Lambda\delta M_{\Lambda N}+\rho_{\Xi^-}\delta M_{\Xi^- N}+\rho_{\Sigma^-}\delta M_{\Sigma^- N}) + {\cal E}_{\rm pot}(\lbrace\rho_i\rbrace) \cr
&-&\Bigg\lbrace\rho_p(\mu_{K^-}+\Sigma_{Kp})+\rho_\Lambda\Sigma_{K\Lambda}+\rho_{\Xi^-}\Big(-\mu_{K^-}+\Sigma_{K\Xi^-}\Big) \cr
&+&\rho_n\Big(\frac{1}{2}\mu_{K^-}+\Sigma_{Kn}\Big)
+\rho_{\Sigma^-}\Big(-\frac{1}{2}\mu_{K^-}+\Sigma_{K\Sigma^-}\Big)\Bigg\rbrace(1-\cos\theta) \cr
&-&\sum_{i=p,\Lambda,\Xi^-,n,\Sigma^-}\mu_i\rho_i \cr
&-&\frac{1}{2}f^2\mu_{K^-}^2\sin^2\theta+f^2m_K^2(1-\cos\theta) \ .
\label{eq:eb}
\end{eqnarray}
The volume part of the energy density of the KCYN after inclusion of the baryon potential energy effects, ${\cal E}_{\rm vol}'(\theta,\lbrace\rho_a \rbrace)$, is given as 
\begin{equation}
 {\cal E}_{\rm vol}'(\theta,\lbrace\rho_a \rbrace)={\cal E}_{\rm kin}(\lbrace\rho_i\rbrace)+{\cal E}_{Y-N{\rm mass}}(\lbrace\rho_i\rbrace)+{\cal E}_{\rm pot}(\lbrace\rho_i\rbrace)+{\cal E}_{K-B{\rm int.}}(\theta,\lbrace\rho_i \rbrace)+{\cal E}_{{\rm free}K}(\theta,\rho_{K^-}) 
 \label{eq:vol-energy}
\end{equation}
with
\begin{subequations}\label{eq:evol}
\begin{eqnarray}
{\cal E}_{\rm kin}(\lbrace\rho_i\rbrace) &=&\frac{3}{5}\frac{(3\pi^2)^{2/3}}{2M_N}\Big(\rho_p^{5/3}+\rho_\Lambda^{5/3} 
 +\rho_{\Xi^-}^{5/3}+\rho_n^{5/3}+\rho_{\Sigma^-}^{5/3}\Big) \ , \label{eq:evolkin} \\
{\cal E}_{Y-N{\rm mass}}(\lbrace\rho_i\rbrace)&=&\rho_\Lambda\delta M_{\Lambda N}+\rho_{\Xi^-}\delta M_{\Xi^- N}+\rho_{\Sigma^-}\delta M_{\Sigma^- N} \ ,  \label{eq:evolynmass} \\
{\cal E}_{K-B{\rm int}}(\theta,\lbrace\rho_i \rbrace)&=&-\sum_{i=p, \Lambda, \Xi^-, \Sigma^-, n}\rho_i\Sigma_{Ki}(1-\cos\theta) \ , \label{eq:evolkb} \\
{\cal E}_{{\rm free}K}(\theta,\rho_{K^-}) 
&=&\frac{1}{2}f^2\mu_{K^-}^2\sin^2\theta+f^2m_K^2(1-\cos\theta) \ . \label{eq:evolfk}
\end{eqnarray}
\end{subequations}
The first term, ${\cal E}_{\rm kin}(\lbrace\rho_i\rbrace)$ , on the right hand side of Eq.~(\ref{eq:vol-energy}) [Eq.~(\ref{eq:evolkin})] denotes the baryon kinetic energy density with the free nucleon mass $M_N$, the second term, ${\cal E}_{Y-N{\rm mass}}(\lbrace\rho_i\rbrace)$ [Eq.~(\ref{eq:evolynmass})], comes from the mass difference between the hyperon and nucleon, $\delta M_{YN}$, the third term, ${\cal E}_{\rm pot}(\lbrace\rho_i\rbrace)$ [Eq.~(\ref{eq:epot})], the baryon potential energy density, the fourth term, ${\cal E}_{K-B{\rm int.}}(\theta,\lbrace\rho_i \rbrace)$ [Eq.~(\ref{eq:evolkb})], the $s$-wave kaon-baryon scalar interaction brought about by the kaon-baryon sigma terms $\Sigma_{Ki}$, and the last term, ${\cal E}_{{\rm free}K}(\theta,\rho_{K^-})$ [Eq.~(\ref{eq:evolfk})], stands for the free part of the condensed kaon energy density (coming from  kinetic energy and free mass terms).

\ \ The kaon condensates obey the  classical field equation. From $\partial{\cal E}_{\rm eff}'(\theta, \lbrace\mu_a\rbrace)/\partial\theta=0$, one obtains
\begin{equation}
\sin\theta\Bigg\lbrack \mu_{K^-}^2\cos\theta-m_K^2+\frac{\mu_{K^-}}{f^2}\Big(\rho_p+\frac{1}{2}\rho_n-\rho_{\Xi^-}-\frac{1}{2}\rho_{\Sigma^-}\Big)+\frac{1}{f^2}\sum_{i=p, \Lambda,\Xi^-, \Sigma^-, n}\rho_i\Sigma_{Ki}\Bigg\rbrack=0 \ . 
\label{eq:theta}
\end{equation}

The chemical equilibrium conditions for the strong processes, $K^-  p\rightleftharpoons \Lambda$,  
$K^-  n\rightleftharpoons \Sigma^-$, $K^-  \Lambda\rightleftharpoons \Xi^-$,  [Eq.~(\ref{eq:chemeq})] are written explicitly as 
\begin{subequations}\label{eq:sequil}
\begin{eqnarray}
\mu_{K^-}+\mu_p &=&\mu_\Lambda \ ,  \label{eq:sequil1} \\
\mu_{K^-} + \mu_n &=&\mu_{\Sigma^-} \ , \label{eq:sequil2} \\
\mu_{K^-} + \mu_\Lambda &=& \mu_{\Xi^-}  \ , \label{eq:sequil3} 
\end{eqnarray} 
\end{subequations}
where the chemical potentials for the baryons are given by $\mu_i=\partial{\cal E}_{\rm vol}'(\theta, \lbrace\rho_a\rbrace)/\partial\rho_i$ with the use of Eqs.~(\ref{eq:rhok}) and (\ref{eq:vol-energy}):
\begin{subequations}\label{eq:mu}
\begin{eqnarray} 
\mu_p& = & \frac{(3\pi^2\rho_p)^{2/3}}{2M_N} -(\mu_{K^-}+\Sigma_{Kp})(1-\cos\theta)+V_p \ , \label{eq:mup} \\
\mu_\Lambda& = & \frac{(3\pi^2\rho_\Lambda)^{2/3}}{2M_N} -\Sigma_{K\Lambda}(1-\cos\theta)
+\delta M_{\Lambda N}+V_\Lambda \ , \label{eq:mul} \\
\mu_{\Xi^-}& = &  \frac{(3\pi^2\rho_{\Xi^-})^{2/3}}{2M_N} -\Big(-\mu_{K^-}+\Sigma_{K\Xi^-}\Big)(1-\cos\theta)
+\delta M_{\Xi^- N}+V_{\Xi^-} \ , \label{eq:muxi} \\
\mu_n& = &\frac{(3\pi^2\rho_n)^{2/3}}{2M_N} -\Big(\frac{1}{2}\mu_{K^-}+\Sigma_{Kn}\Big)(1-\cos\theta)+V_n \ ,  \label{eq:mun} \\
\mu_{\Sigma^-}& = &  \frac{(3\pi^2\rho_{\Sigma^-})^{2/3}}{2M_N} -\Big(-\frac{1}{2}\mu_{K^-}+\Sigma_{K\Sigma^-}\Big)(1-\cos\theta)
+\delta M_{\Sigma^- N}+V_{\Sigma^-} \ . \label{eq:mus} 
\end{eqnarray}
\end{subequations}

By solving the classical kaon field equation (\ref{eq:theta}) together with the three equations (\ref{eq:conv1})$-$(\ref{eq:conv3}) for charge, strangeness, and baryon number conservations, and the three chemical equilibrium conditions for the strong processes, (\ref{eq:sequil1})$-$(\ref{eq:sequil3}) at a given baryon number density $\rho_{\rm B}$, one obtains $\mu_{K^-}$, $\theta$, and the number densities of baryons. With these quantities, the volume part of the energy density, ${\cal E}_{\rm vol}'$ [(\ref{eq:vol-energy})], or the total energy, $E/A$ [(\ref{eq:energy})], for the ground state of the KCYN is obtained.

\section{Structure of the KCYN}
\label{sec:4}

\ \ In this section, we present the numerical results on the energy of the nucleus with kaon condensates and with mixing of hyperons in the case of $A$ = 20, $Z$ = 10, and $|S|$ = 16 as an example, and  discuss the ground state properties of the KCYN such as equilibrium density, particle fractions, and binding energy of the nucleus. 

\begin{figure}[h]
\noindent\begin{minipage}[l]{0.50\textwidth}
\begin{center}
\includegraphics[height=.3\textheight]{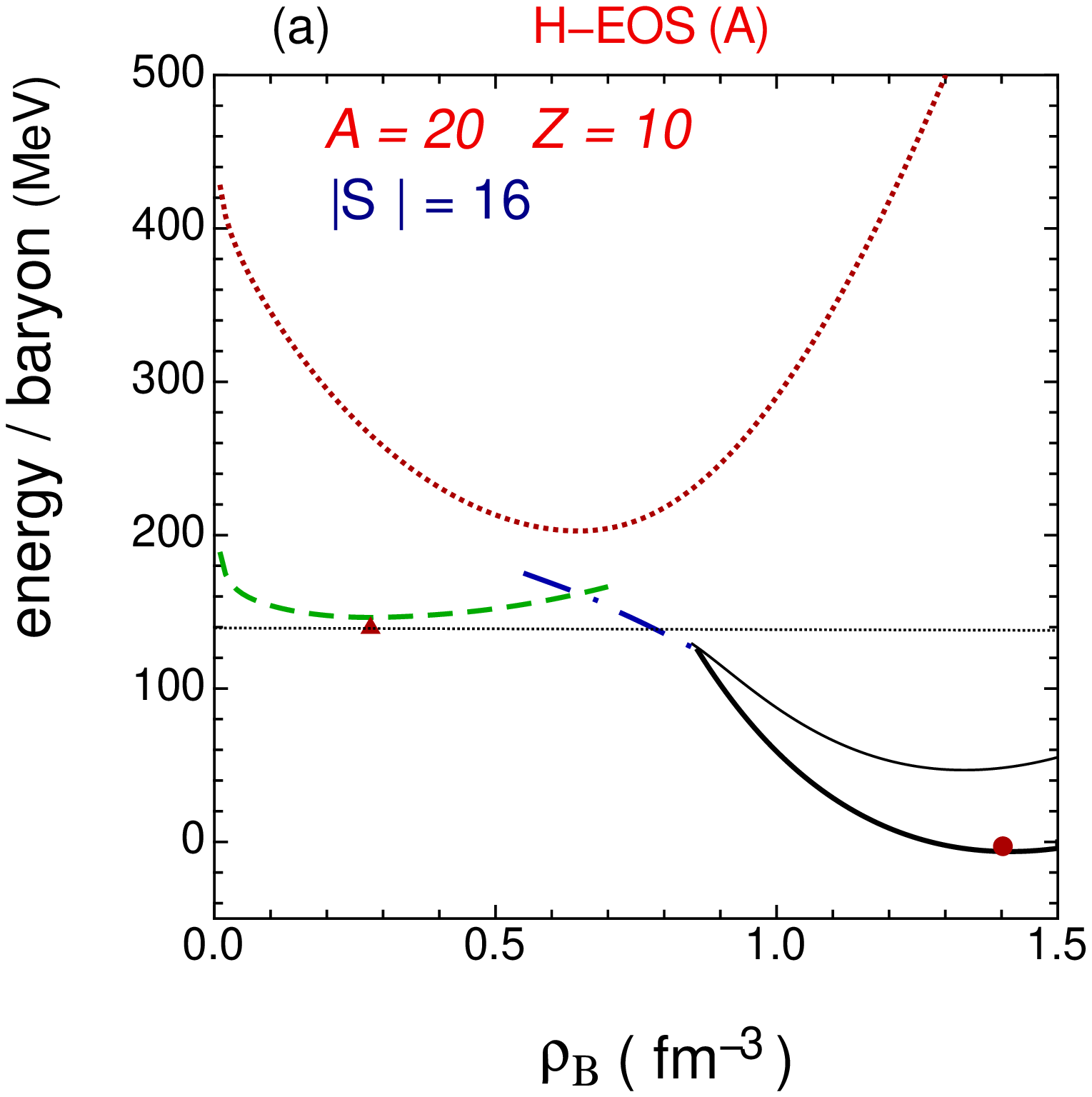}
\end{center}
\end{minipage}~
\begin{minipage}[r]{0.50\textwidth}
\begin{center}
\includegraphics[height=.3\textheight]{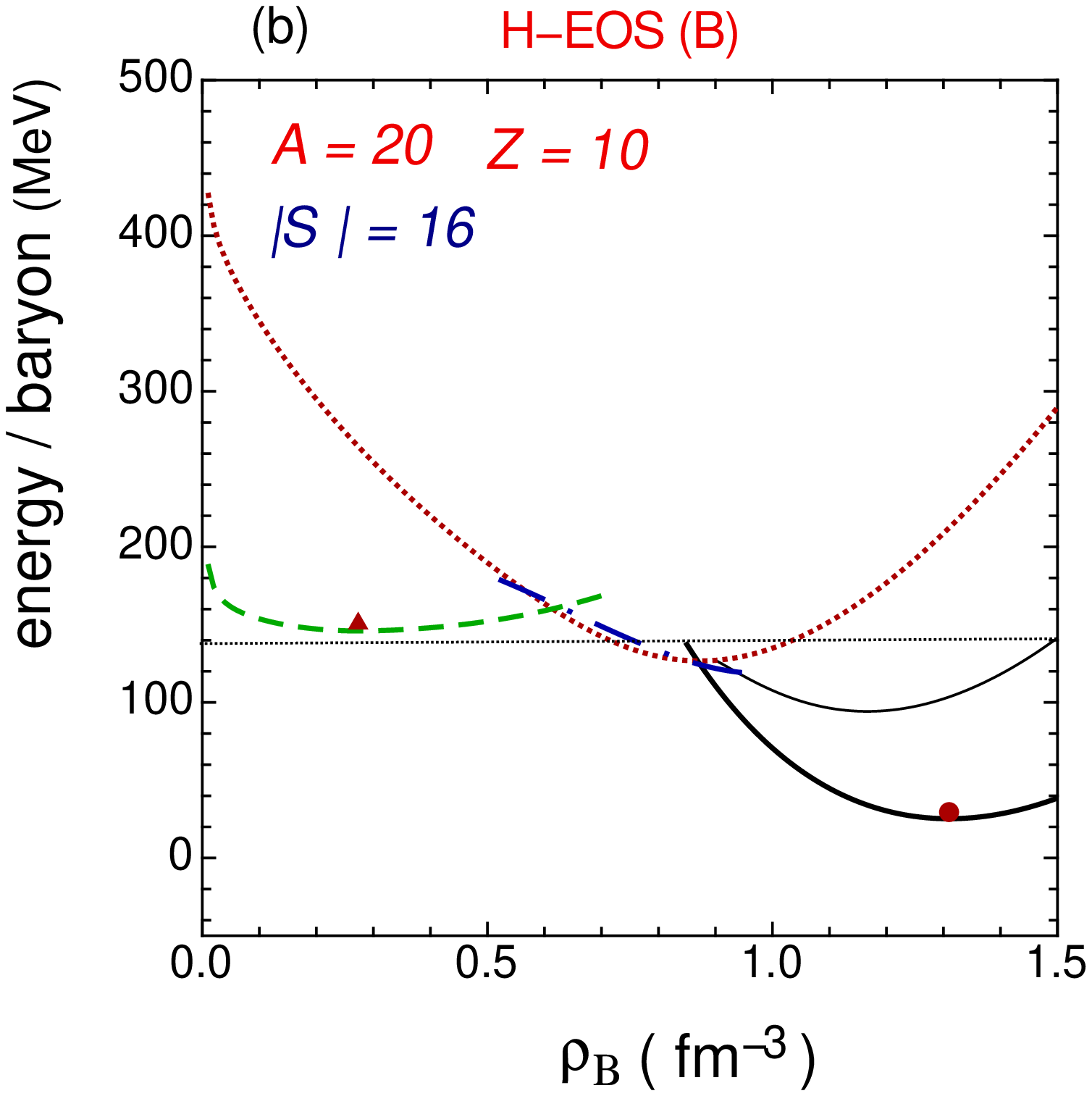}
\end{center}
\end{minipage}
\caption{(a) The energy per baryon of the nucleus as functions of the baryon number density inside the nucleus $\rho_{\rm B}$ for H-EOS (A) and $\Sigma_{Kn}$=305 MeV. The case for $A$=20, $Z$=10, and $|S|$=16 is shown. The bold solid line is for the KCYN with baryons ($p$, $\Lambda$, $\Xi^-$, $n$, $\Sigma^-$), the dash-dotted line for the KCYN with baryons ($p$, $\Lambda$, $n$), and the dashed line is for the noncondensed hypernucleus. For comparison, the energy per baryon for the kaon-condensed nucleus with nucleons only ($p$, $n$) is shown by the dotted line. Also the one for the KCYN with baryons ($p$, $\Lambda$, $n$, $\Sigma^-$) without allowing for mixing of the $\Xi^-$ is inserted by the thin solid line. (b) The same as in (a), but for H-EOS (B).}%
\label{fig1}
\end{figure}

\subsection{Ground state energy}
\label{subsec:4-1}

\ \ The energy per baryon of the KCYN, $E(\rho_{\rm B}; A,Z-|S|, |S|)/A$, with $A$=20, $Z$=10, and $|S|$=16 are shown as functions of the baryon number density inside the nucleus, $\rho_{\rm B}$, in Fig.~\ref{fig1}. The H-EOS (A) and H-EOS (B) are used as the baryon-baryon interaction model in Figs.~\ref{fig1}(a) and (b), respectively. 
From Fig.~\ref{fig1}~(a), one finds that 
the ground state shifts from the noncondensed hypernuclear state (dashed line) at lower densities to the $K^-$-condensed state with baryons ($p$, $\Lambda$, $n$) and without mixing of $\Xi^-$ and $\Sigma^-$ (dash-dotted line) as $\rho_{\rm B}$ increases, and to the $K^-$-condensed state with baryons ($p$, $\Lambda$, $\Xi^-$, $n$, $\Sigma^-$) (bold solid line) at higher densities.
 Figs.~\ref{fig1}~(a) and (b) show qualitatively similar features between H-EOS~(A) and H-EOS~(B) for transition of the ground state with increase in density $\rho_{\rm B}$. 
It is to be noted that these transitions of the ground states occur discontinuously as density $\rho_{\rm B}$ increases. 

For both H-EOS~(A) and H-EOS~(B), 
there are two energy minima with respect to $\rho_{\rm B}$. One is for the kaon-condensed hypernucleus with baryons ($p$, $\Lambda$, $\Xi^-$, $n$, $\Sigma^-$) (denoted as a dot on the bold solid line), and the other is for the noncondensed hypernucleus composed of ($\Lambda$, $\Xi^-$, $n$) (denoted as a triangle on the dashed line). [With regard to the particle composition for these minimum states, see Figs.~\ref{fig6} and \ref{fig7} in Sec.~\ref{subsubsec:5-2-2}.] 
Properties of the KCYN at the equilibrium density $\rho_{\rm B}^0$ for $A$=20, $Z$=10, and $|S|$=16 are listed in Table~\ref{tab:profile}. 
\begin{table}[!]
\caption{Properties of the KCYN at the equilibrium state for $A$=20, $Z$=10, and $|S|$=16. The H-EOS (A) and (B) stand for the baryon-baryon interaction models. $R_A$ is the radius of the KCYN, $\rho_{\rm B}^0$ the baryon number density inside the nucleus, $\rho_a/\rho_{\rm B}^0$ 
($a$ = $K^-, p, \Lambda, \Xi^-, n, \Sigma^-$) the particle fractions, $E_{\rm B}/A$ the binding energy per baryon, and $\Delta E/A$ is the difference of the energy per baryon from that of the free $\Lambda$-nucleon system with the same $A$, $Z$, and $|S|$ ($A$=20, $Z$=10, and $|S|$=16).  }
\label{tab:profile}
\begin{center}
\begin{tabular}{c||c|c|c|c|c|c|c|c|c|c}
\hline
H-EOS & $R_A$ & $\rho_{\rm B}^0$ & $\rho_{K^-}/\rho_{\rm B}^0$ & $\rho_p/\rho_{\rm B}^0$ & $\rho_\Lambda/\rho_{\rm B}^0$ & $\rho_{\Xi^-}/\rho_{\rm B}^0$ & $\rho_n/\rho_{\rm B}^0$ & $\rho_{\Sigma^-}/\rho_{\rm B}^0$ & $-E_{\rm B}/A$ & $\Delta E/A$ \\
      & (fm) & (fm$^{-3}$) & & & & & & & (MeV)&(MeV) \\\hline
(A)   & 1.50 & 1.41 & 0.10 & 0.11 & 0.26 & 0.13 & 0.31 & 0.19 & $-$393 & $-$147 \\
(B)   & 1.54 & 1.31 & 0.10 & 0.12 & 0.23 & 0.15 & 0.32 & 0.18 & $-$361 & $-$116 \\\hline
\end{tabular}
\end{center}
\end{table}
The equilibrium density $\rho_{\rm B}^0$ for the KCYN is read as 1.41 fm$^{-3}(=$8.8 $\rho_0$) for H-EOS~(A) and 1.31 fm$^{-3}$(=8.2 $\rho_0$) for H-EOS~(B). 
The radius $R_A$ is given by 
$\displaystyle R_A=\lbrack3A/(4\pi\rho_{\rm B}^0)\rbrack^{1/3}$$\simeq$ 0.6 $A^{1/3}$ fm, which should be compared with the radius of the normal nucleus, $\simeq$ 1.2 $A^{1/3}$ fm. Thus the KCYN with the equilibrium density $\rho_{\rm B}^0$ and high strangeness fraction, $|S|/A$=$O(1)$, is a highly dense and compact object.

In Table~\ref{tab:profile}, the binding energy per baryon, $E_{\rm B}/A$, is defined by 
 $E_{\rm B}/A$=$-\lbrack E_0(A, Z-|S|, |S|)-(|S|m_K+E_0(A, Z, 0))\rbrack/A$, where $E_0(A, Z-|S|, |S|)$ is the ground state energy of the nucleus with $A$, $Z$, and $|S|$ at the equilibrium density.
$\Delta E/A$ (=$E_0(A, Z-|S|, |S|)/A-\delta M_{\Lambda N}\cdot |S|/A$) is the difference of the energy per baryon from that of the free $\Lambda$-nucleon system with the same $A$, $Z$, and $|S|$ ($A$=20, $Z$=10, and $|S|$=16).
One finds that the energy of the KCYN at $\rho_{\rm B}^0$  is even lower than that of the free $\Lambda$-nucleon system where the energy per baryon is given by $\displaystyle |S|/A\cdot \delta M_{\Lambda N}\sim$ 140 MeV (shown by the thin straight line in parallel with the $\rho_{\rm B}$ axis in Figs.~\ref{fig1}~(a) and (b)). Therefore, 
it may be stable against the strong processes and is expected to have a long lifetime, decaying only through weak processes. 
The binding energy and saturation density at the energy minimum of the KCYN in the case of H-EOS (B) are smaller than those in the case of H-EOS (A), which stems from the fact that the repulsive interactions between hyperons and nucleons from many-body terms at high densities become more dominant for H-EOS (B) than for H-EOS (A)\cite{m07}. 

As listed in Table~\ref{tab:profile}, part of the strangeness $|S|$ is carried by hyperons as well as kaon condensates, mainly by the $\Lambda$, but also by heavier hyperons $\Sigma^-$ and $\Xi^-$ since the equilibrium density $\rho_{\rm B}^0$ is high. For reference, we show the $K^-$-condensed state with baryons ($p$, $\Lambda$, $n$, $\Sigma^-$) without allowing for mixing of the $\Xi^-$ is inserted by the thin solid line\cite{m07-2} in Figs.~\ref{fig1}~(a) and (b).
 One can see that mixing of $\Xi^-$ further leads to deeper binding of the KCYN with higher equilibrium density. It is to be noted that the chiral angle $\theta$ for the KCYN with with baryons ($p$, $\Lambda$, $\Xi^-$, $n$, $\Sigma^-$) is $\pi$ (rad), corresponding to the limiting case of the fully-developed condensates. [see Sec.~\ref{subsubsec:5-2-1} for details.]
 
As a scale of the stiffness of the KCYN, we estimate the incompressibility $K$ 
at $\rho_{\rm B}^0$, which is defined by $K\equiv 9\rho_{\rm B}^2\partial^2 (E/A)/\partial\rho_{\rm B}^2\vert_{\rho_{\rm B}=\rho_{\rm B}^0} $. For $A$ = 20, one obtains $K\sim 3\times 10^3$ MeV, which is larger by an order of magnitude than the standard value 210 MeV 
for the normal nuclear matter\cite{b80}. 
The measurement of an extremely small radius and/or huge incompressibility for the nucleus with a given $A$ would give us a clue to find a highly dense and compact object. 

With regard to the local energy minimum for the noncondensed hypernucleus, the equilibrium density $\rho_{\rm B}^0$ is 
read from Figs.~\ref{fig1}~(a) and (b) as $\simeq$ 0.28 fm$^{-3}$ ($\sim$ 1.8$\rho_0$) for both H-EOS~(A) and H-EOS~(B). This hypernucleus is a self-bound system, but the energy per baryon marginally exceeds that of the free $\Lambda$ and nucleon system for the same $A$, $Z$, and $|S|$.

\subsection{Effects of hyperon-mixing}
\label{subsec:4-2}

\ \ Here we discuss mechanisms leading to 
a deeply bound and highly dense state of the KCYN in the case of a large fraction of the strangeness $|S|/A$ = $O(1)$ as discussed in Sec.~\ref{subsec:4-1}. In Figs.~\ref{fig1}(a) and (b), we show the energy per baryon of the $K^-$-condensed state without including hyperons by putting $\rho_\Lambda=\rho_{\Xi^-}=\rho_{\Sigma^-}$ = 0 as a function of $\rho_{\rm B}$ by the dotted line in addition to the curves of the ground state energy shown by the dashed, dash-dotted and solid lines. (In this paper, we abbreviate the kaon-condensed nuclei without mixing of hyperons as the KCN and distinguish it from the KCYN.) In this case,  the whole strangeness $|S|$ is carried solely by the $K^-$ condensates, so that one has a kaon-condensed state at any density.  
From comparison of the dotted line and bold solid line, 
one finds that mixing of hyperons leads to a more deeply bound and denser equilibrium state of the KCYN.

\begin{figure}[!]
\noindent\begin{minipage}[l]{0.50\textwidth}
\begin{center}
\includegraphics[height=.3\textheight]{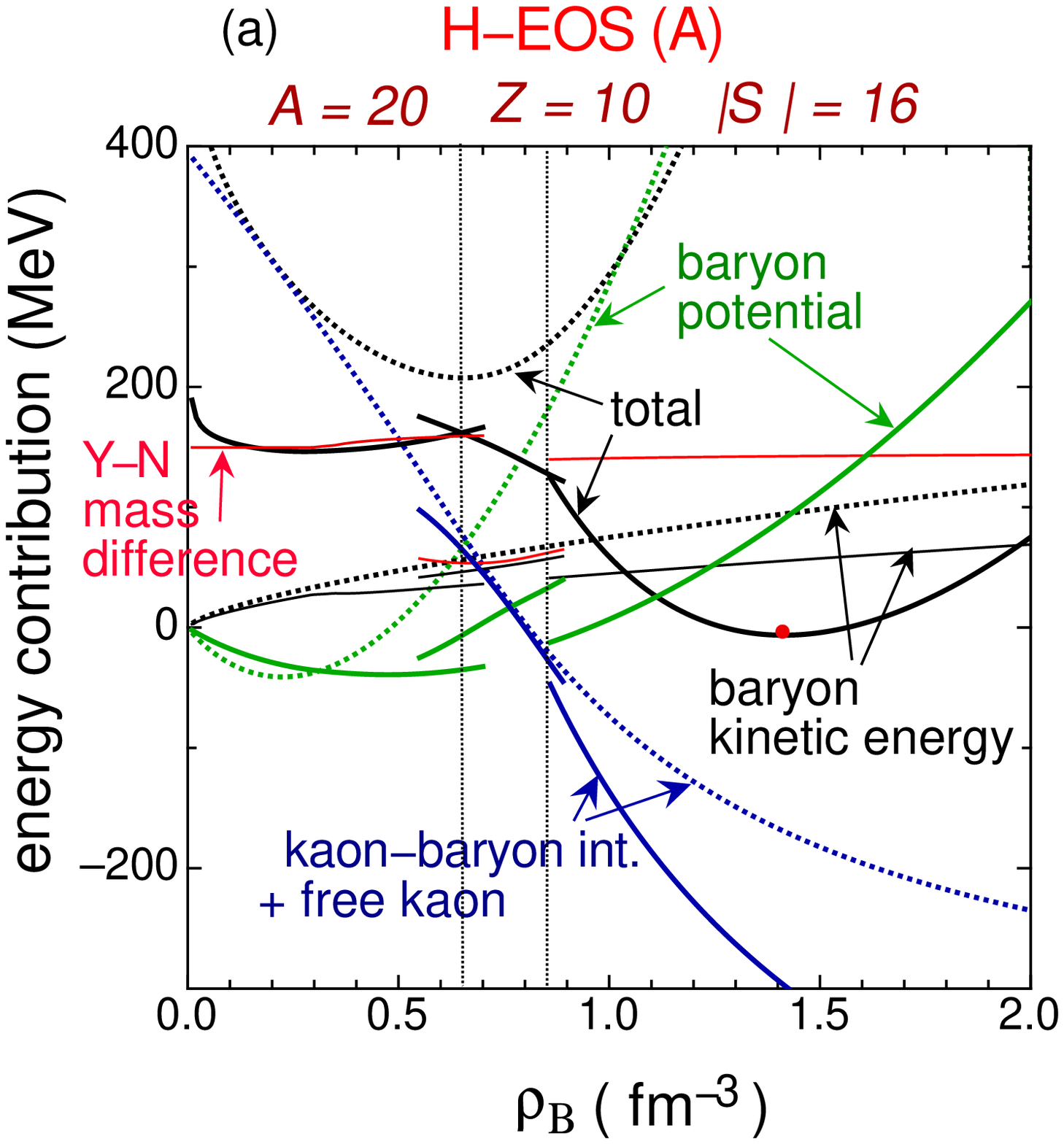}
\end{center}
\end{minipage}~
\begin{minipage}[r]{0.50\textwidth}
\begin{center}
\includegraphics[height=.3\textheight]{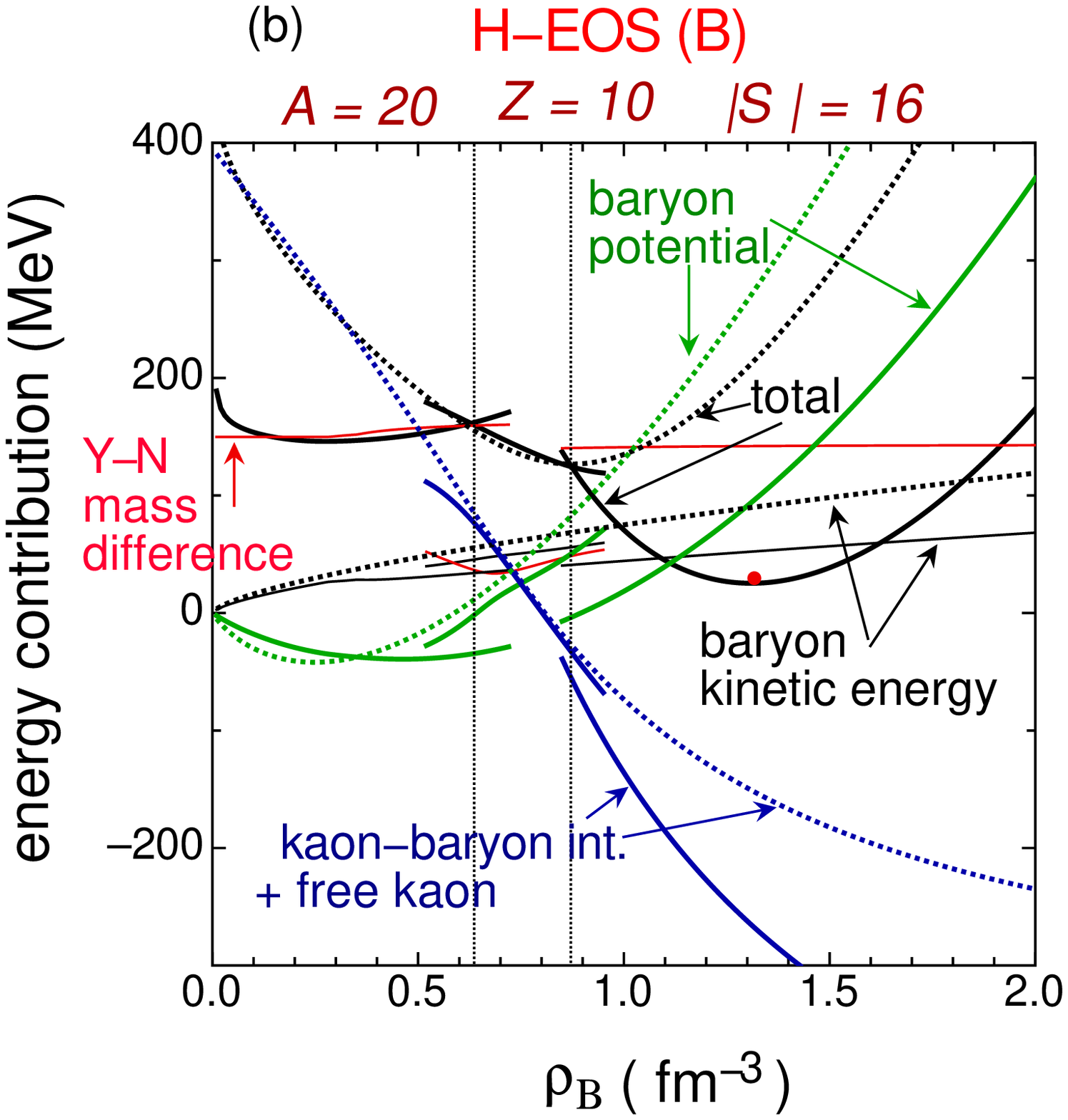}
\end{center}
\end{minipage}
  \caption{(a) The contributions to the total energy per baryon for the ground state of the nucleus with $A$=20, $Z$=10, and $|S|$=16  (solid lines) and those for the kaon-condensed nucleus with nucleons only (dotted lines) as functions of $\rho_{\rm B}$ for H-EOS (A) and $\Sigma_{Kn}$=305 MeV. (b) The same as in (a), but for H-EOS (B).}
\label{fig2}
\end{figure}

In Fig.~\ref{fig2}, the contributions to the total energy per baryon of the ground state with and without mixing of hyperons (the solid lines and dotted lines, respectively) are shown as functions of $\rho_{\rm B}$ for $A$=20, $Z$=10, and $|S|$=16. The result by the use of H-EOS (A) [H-EOS~(B)] is shown in Fig.~\ref{fig2}~(a) [Fig.~\ref{fig2}~(b)].
The dependence of the total energy per baryon on $\rho_{\rm B}$ is mainly determined from the two contributions: (I) the sum of the $s$-wave kaon-baryon interaction and free kaon energy, ${\cal E}_{K-B{\rm int}}/\rho_{\rm B}+{\cal E}_{{\rm free}K}/\rho_{\rm B}$ [with Eqs.~(\ref{eq:evolkb}) and (\ref{eq:evolfk})], and (II) the baryon potential energy ${\cal E}_{\rm pot}/\rho_{\rm B}$ [with Eq.~(\ref{eq:epot})]. The energy minimum appears as a result of balance of these two energy contributions. 
In the density region  $\rho_{\rm B}\gtrsim 0.86$ fm$^{-3}$ for H-EOS~(A) [$\rho_{\rm B}\gtrsim 0.88$ fm$^{-3}$ for H-EOS~(B)], 
the attractive effect from (I), coming from ${\cal E}_{K-B{\rm int}}/\rho_{\rm B}$, gets more remarkable for the hyperon-mixing case than for the case without hyperons at high densities. On the other hand, 
the repulsive effect from (II), which contributes to increase the energy as density increases, is much reduced for the hyperon-mixing case, since the nucleon-nucleon repulsion is avoided by mixing of hyperons at high densities\cite{nyt02}. These hyperon-mixing effects lead the energy minimum state of the KCYN to a deeply bound and highly dense object as compared with the case of the KCN. 

We comment on the behaviors of the other contributions to the total energy per baryon. The baryon kinetic energy, ${\cal E}_{\rm kin}/\rho_{\rm B}$ [with Eq.~(\ref{eq:evolkin})], depends on $\rho_{\rm B}$ only weakly. Mixing of hyperons leads to reduce the ${\cal E}_{\rm kin}/\rho_{\rm B}$ from the case without mixing of hyperons, but its effect is not marked. The $Y-N$ mass difference term, ${\cal E}_{Y-N {\rm mass}}/\rho_{\rm B} $ [with Eq.~(\ref{eq:evolynmass})], pushes up the total energy per baryon, but its $\rho_{\rm B}$-dependence is negligible in each density region of the noncondensed hypernucleus, KCYN with baryons ($p$, $\Lambda$, $n$), and KCYN with baryons ($p$, $\Lambda$, $\Xi^-$, $n$, $\Sigma^-$). The numerical results show that the surface energy term, ${\cal E}_{\rm surface}/\rho_{\rm B}$ (= $\sigma\lbrack 36\pi/(A\rho_{\rm B}^2)\rbrack^{1/3}$), and the Coulomb energy term, ${\cal E}_{\rm Coulomb}/\rho_{\rm B}$ (= $\displaystyle \frac{3}{5}(Z-|S|)^2e^2\lbrack4\pi\rho_{\rm B}/(3A^4)\rbrack^{1/3}$), are very small as compared with the other energy contributions: (${\cal E}_{\rm surface}/\rho_{\rm B}$, ${\cal E}_{\rm Coulomb}/\rho_{\rm B}$) = (38 MeV, 0.20 MeV)$\rightarrow$(1.0 MeV, 1.3 MeV) as $\rho_{\rm B}$ = ($\sim$ 0 $\rightarrow$ 2.0) fm$^{-3}$ for both cases of H-EOS~(A) and (B). Hence these two terms are not shown in Fig.~\ref{fig2}. 
The negligible contribution of the surface energy term to the total energy indicates that the energy and static properties of the KCYN are mostly determined by the particle fractions, $\rho_a/\rho_{\rm B}^0$ ($a$ = $K^-$, $p$, $\Lambda$, $\Xi^-$, $n$, $\Sigma^-$) and strangeness and electric charge fractions, $|S|/A$, $(Z-|S|)/A$, not the particle numbers, and $A$, $Z$, $|S|$ themselves.

\section{Systematic change of the structure of the KCYN}
\label{sec:5}

\ \ We consider the properties of the energy minimum state of the KCYN systematically by changing the strangeness number $|S|$, while keeping the baryon number and atomic number fixed as $A$ = 20 and $Z$ = 10.

\begin{figure}[h]
\noindent\begin{minipage}[l]{0.50\textwidth}
\begin{center}
\includegraphics[height=.3\textheight]{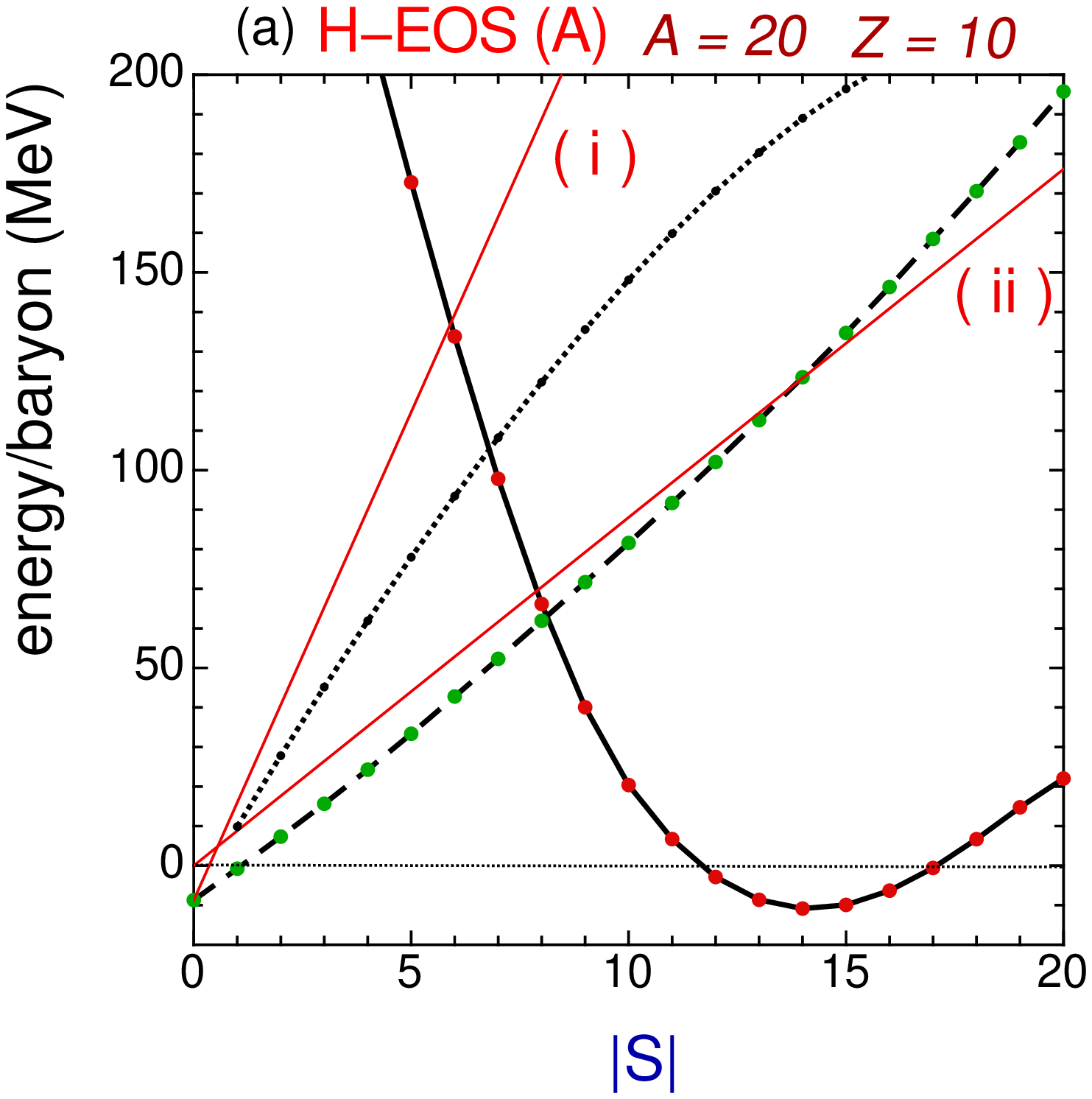}
\end{center}
\end{minipage}~
\begin{minipage}[r]{0.50\textwidth}
\begin{center}
\includegraphics[height=.3\textheight]{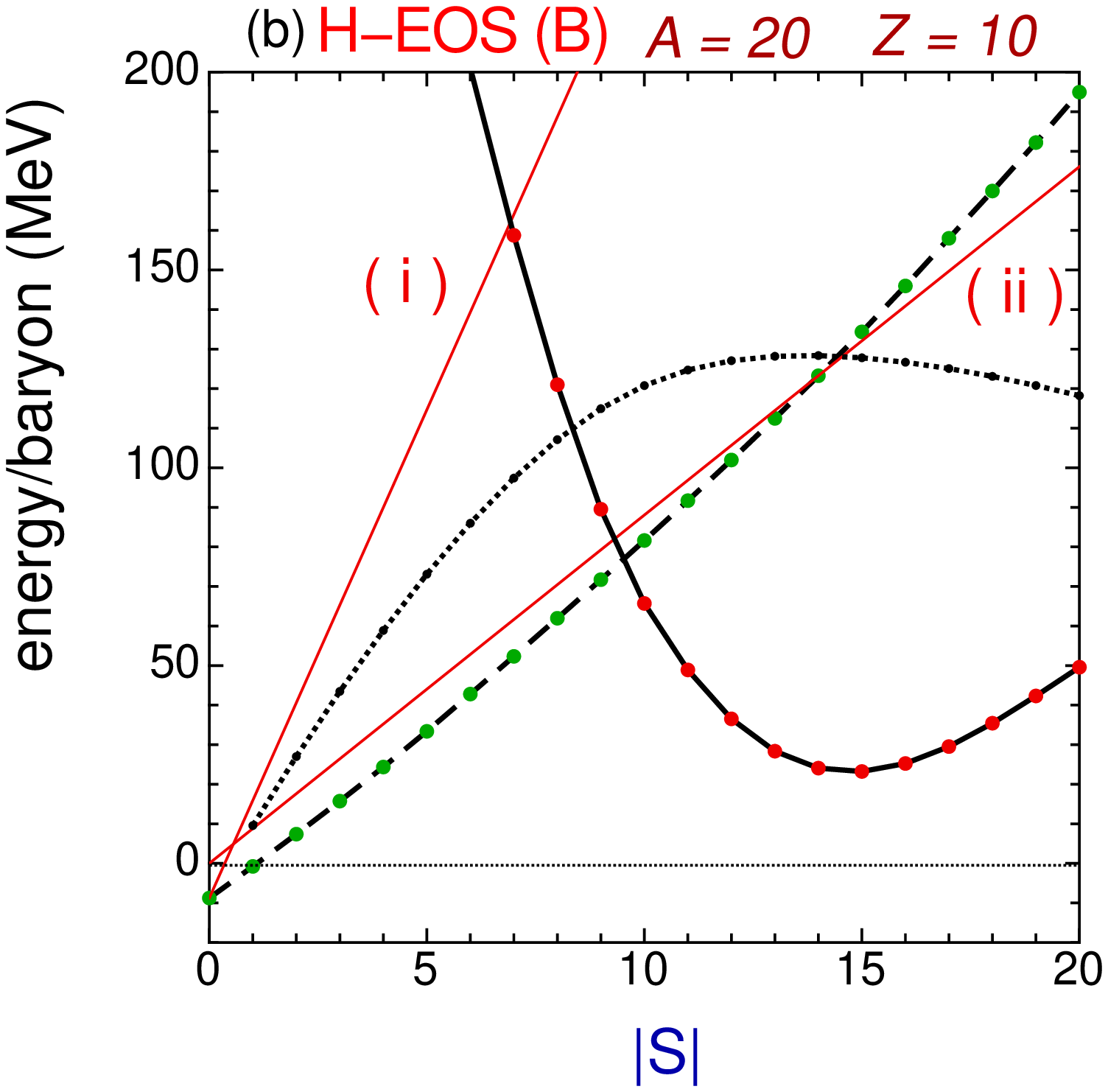}
\end{center}
\end{minipage}  
\caption{(a) The energy per baryon for the energy minimum state of the KCYN (the solid line) and that of the noncondensed hypernucleus (the dashed line) as functions of the strangeness number $|S|$ in the case of $A$=20, $Z$=10 and for H-EOS (A) and $\Sigma_{Kn}$=305 MeV. For comparison, the dependence on $|S|$ of the energy per baryon for the energy minimum state of the KCN is depicted  by the dotted line. The straight line (i) stands for the energy per baryon of the system consisting of the free $K^-$ the number of which is $|S|$ and normal nucleus, $(|S|m_K+E_0(A, Z, |S|=0))/A$ with $A$ =20 and $Z$=10. The straight line (ii) represents the energy per baryon of 
the free $\Lambda$-nucleon system where the energy per baryon is given by $\displaystyle |S|/A\cdot \delta M_{\Lambda N}$. (b) The same as in (a), but for H-EOS (B).}
\label{fig3}
\end{figure}

\subsection{Dependence of the ground state energy on $|S|$}
\label{subsec:5-1}

\ \ In Figs.~\ref{fig3}(a) and (b), the energy per baryon for the energy minimum state of the KCYN (the solid line) and that of the noncondensed hypernucleus (the dashed line) are shown as functions of the strangeness number $|S|$ in the case of $A$=20, $Z$=10 for $\Sigma_{Kn}$=305 MeV.
Fig.~\ref{fig3}~(a) is for H-EOS~(A) and (b) is for H-EOS~(B). 
For comparison, the dependence on $|S|$ of the energy per baryon for the energy minimum state of the kaon-condensed nucleus without allowing for mixing of hyperons (KCN) by putting $\rho_{\Lambda}$ = $\rho_{\Xi^-}$ = $\rho_{\Sigma^-}$ = 0 is depicted by the dotted line. The straight line (i) stands for the energy per baryon of the system consisting of the free $K^-$ the number of which is $|S|$ and normal nucleus, $(|S|m_K+E_0(A, Z, |S|=0))/A$. The straight line (ii) represents the energy per baryon of the free $\Lambda$-nucleon system given by $\displaystyle |S|/A\cdot \delta M_{\Lambda N}$.  
From Fig.~\ref{fig3}~(a), one finds that the energy per baryon for the energy minimum state of the KCYN decreases as the strangeness number $|S|$ increases until $|S|$ = 14, where the energy minimum state is most deeply bound with the binding energy per strangeness, $E_{\rm B}/|S|$ = 497 MeV. For $|S|\geq 15$, the energy per baryon turns to increase monotonically with $|S|$. For a sizable amount of strangeness, $|S|\geq 8$, the energy per baryon for the energy minimum state is lower than that for the free $\Lambda$-nucleon system (the straight line (ii)), which implies that the KCYN is stable against the strong decay processes.
For a small amount of the strangeness, $|S|\leq 5$, the energy per baryon exceeds that for the free $K^-$ and normal nucleus (the straight line (i)), so that the KCYN is unbound for such a small value of $|S|$. 

On the other hand, the noncondensed hypernucleus is bound and stable against the strong decay processes even for the small $|S|$ [i.e., the dashed line is below the line (ii)],  while the energy per baryon for the noncondensed hypernucleus increases monotonically with increase in $|S|$, as is seen from Fig.~\ref{fig3}~(a).  
One also finds from Fig.~\ref{fig3}~(a) that the energy per baryon for the energy minimum state of the KCN (the dotted line) lies between the lines (i) and (ii) for the relevant values of $|S|$. Therefore the kaon-condensed nucleus without mixing of hyperons may be a bound state, but it decays through the strong processes, which is consistent with the results by Gazda et al.\cite{gfgm07}. 

The above results for the case of H-EOS~(A) in Fig.~\ref{fig3}~(a) are qualitatively the same as those for the case of H-EOS~(B) in Fig.~\ref{fig3}~(b), except for the $|S|$-dependence of the energy per baryon for the energy minimum state of the KCN (the dotted lines): For H-EOS~(B), the energy per baryon of the KCN has a maximum at $|S|$ = 14, and it turns to decrease for $|S|\geq 15$, where the curve for the energy per baryon lies below the line (ii).

It is concluded that a deeply bound nucleus with kaon condensates which is stable and decays only through weak processes should be formed with the large strangeness fraction $|S|/A\gtrsim 0.5$ and that part of the strangeness $|S|$ should be taken over by hyperons through mixing of hyperons in the nucleus.  

\subsection{Dependence of internal structure on $|S|$}
\label{subsec:5-2}
\begin{figure}[h]
\noindent\begin{minipage}[l]{0.50\textwidth}
\begin{center}
\includegraphics[height=.3\textheight]{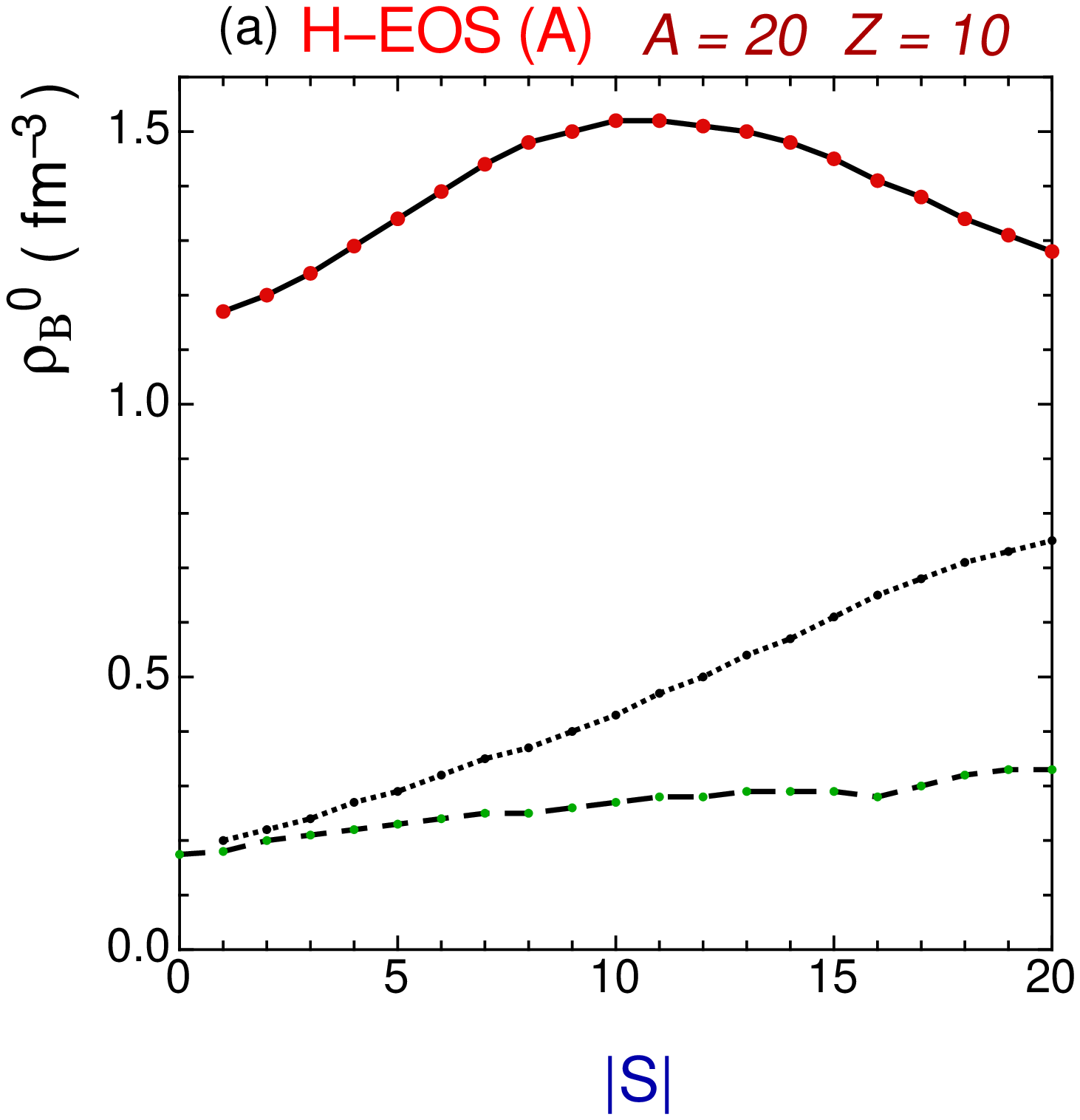}
\end{center}
\end{minipage}~
\begin{minipage}[r]{0.50\textwidth}
\begin{center}
\includegraphics[height=.3\textheight]{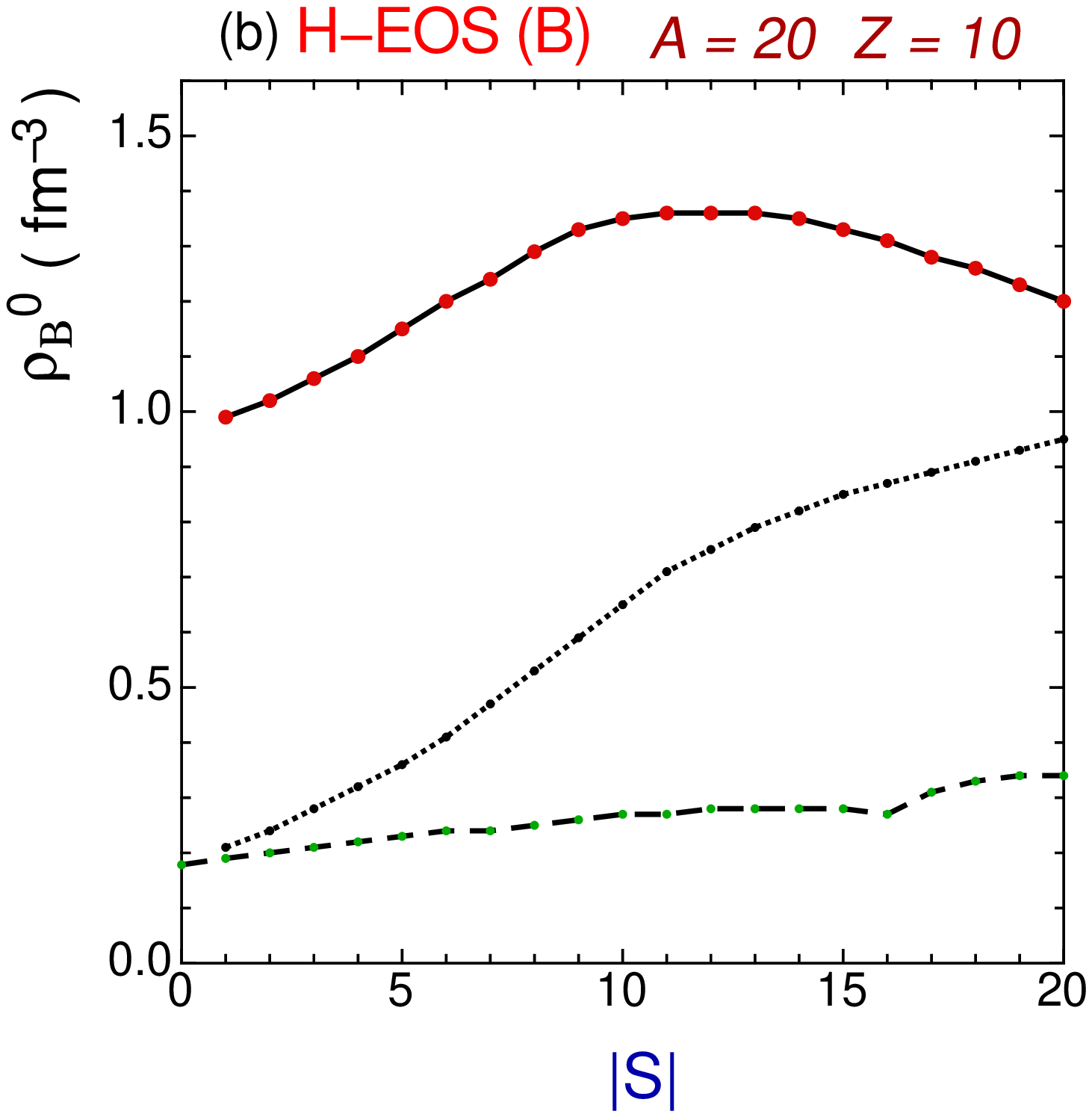}
\end{center}
\end{minipage}
\caption{(a) The baryon number densities $\rho_{\rm B}^0$ inside the nucleus at the energy minimum states of the KCYN (the solid line), the noncondensed hypernucleus (the dashed line), and the KCN (the dotted line), with $A$=20, $Z$=10 as functions of the strangeness $|S|$ for H-EOS (A) and $\Sigma_{Kn}$=305 MeV. (b) The same as in (a), but for H-EOS (B).}
\label{fig4}
\end{figure}

\begin{figure}[h]
\noindent\begin{minipage}[l]{0.50\textwidth}
\begin{center}
\includegraphics[height=.3\textheight]{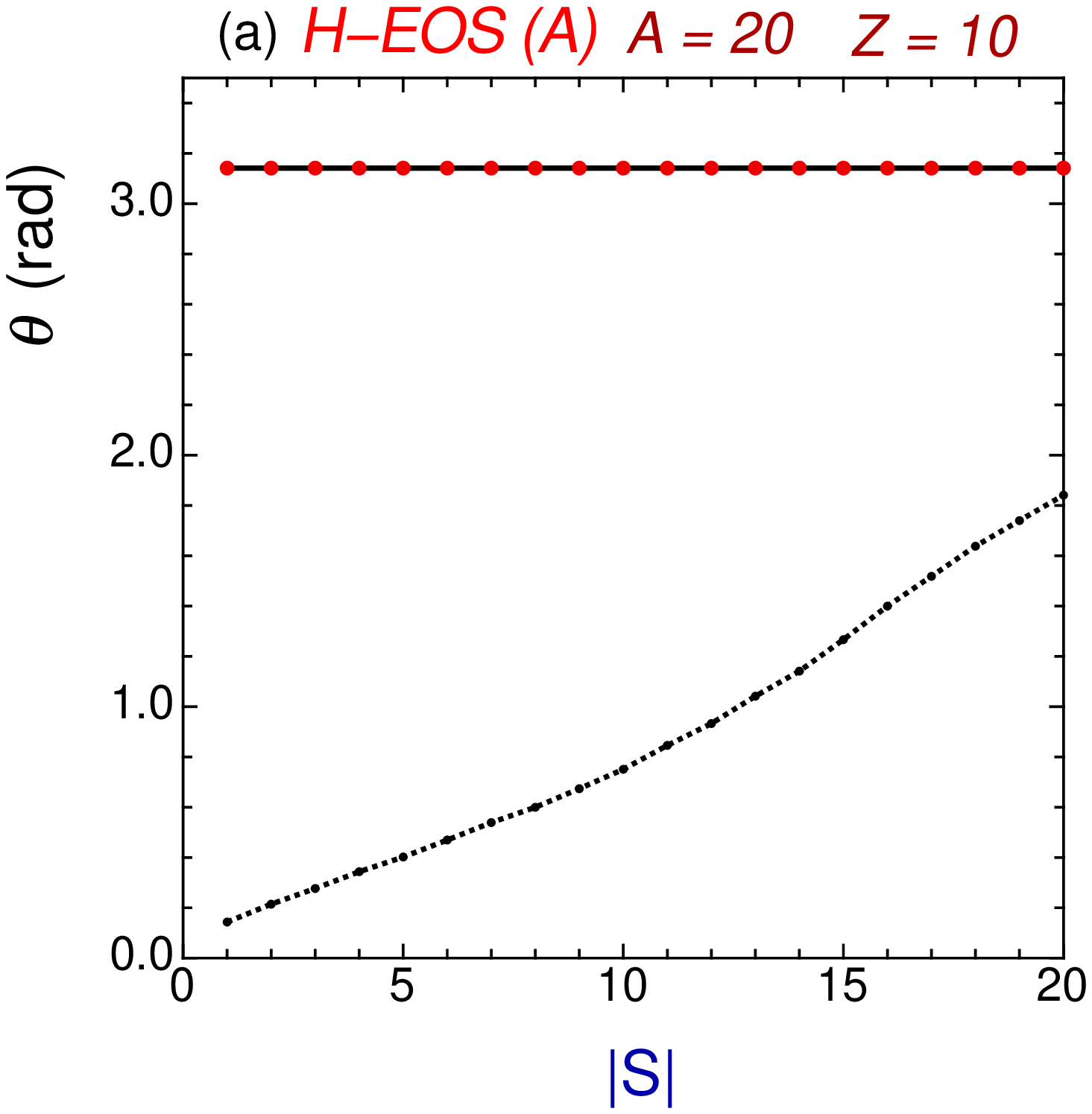}
\end{center}
\end{minipage}~
\begin{minipage}[r]{0.50\textwidth}
\begin{center}
\includegraphics[height=.3\textheight]{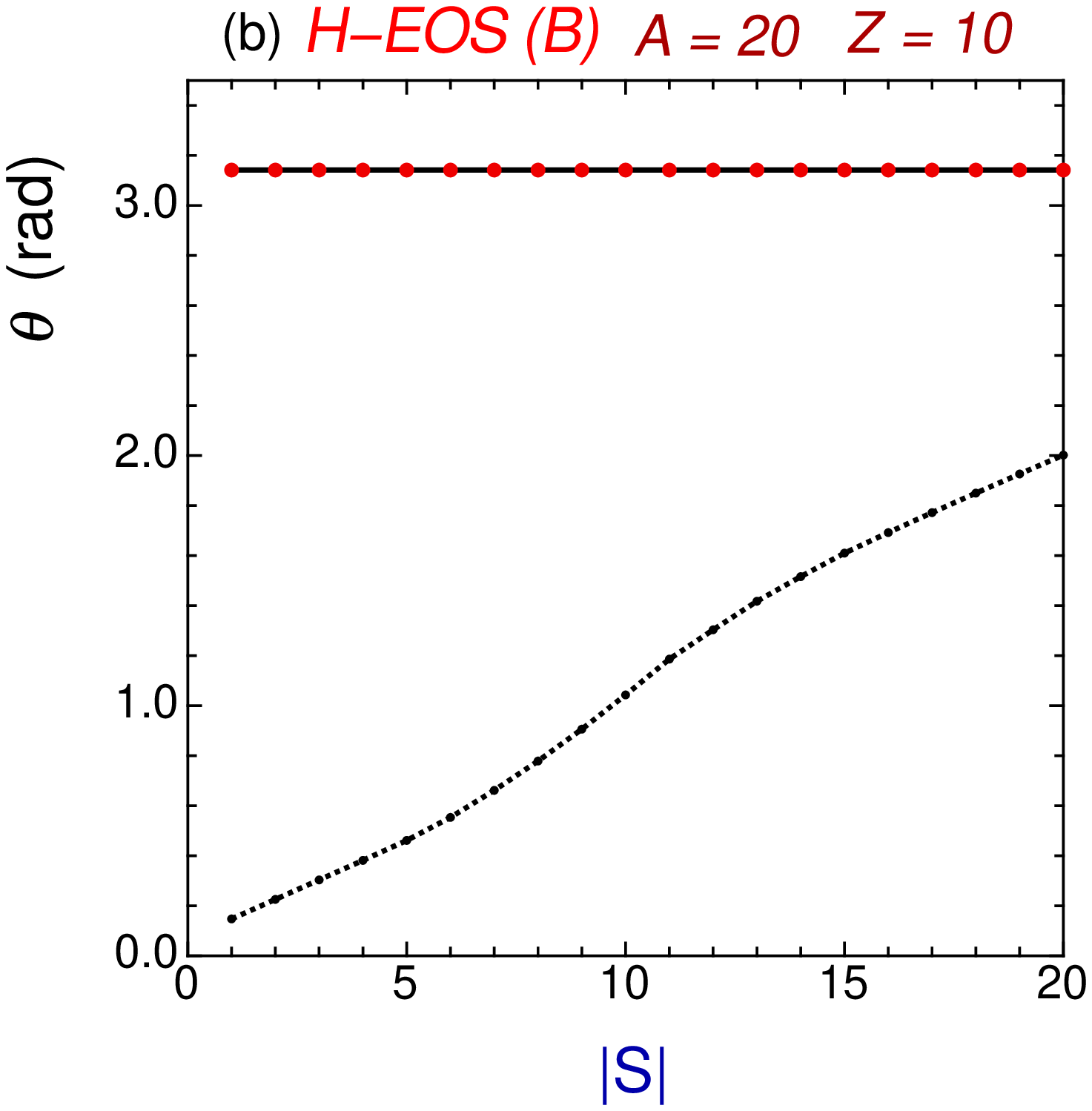}
\end{center}
\end{minipage}
\caption{(a) The chiral angle $\theta$ at the energy minimum state of the KCYN (the KCN) with $A$=20, $Z$=10 as functions of the strangeness $|S|$ for H-EOS (A) and $\Sigma_{Kn}$=305 MeV by the solid line (the dotted line).  (b) The same as in (a), but for H-EOS (B).}
\label{fig5}
\end{figure}

\subsubsection{Baryon number density and strength of kaon condensates inside the nucleus}
\label{subsubsec:5-2-1}

\ \ In Fig.~\ref{fig4}, we show the baryon number densities $\rho_{\rm B}^0$ inside the nucleus at the energy minimum states of the KCYN (the solid line), the noncondensed hypernucleus (the dashed line), and the KCN (the dotted line) with $A$=20, $Z$=10 as functions of the strangeness $|S|$. 
In Fig.~\ref{fig5}, we also show the chiral angle $\theta$ for the classical kaon field inside the nucleus at the energy minimum states of the KCYN (the solid line) and the KCN (the dotted line) with $A$=20, $Z$=10 as functions of the strangeness $|S|$. For each of these figures, (a) is for H-EOS~(A), and (b) is for H-EOS~(B). 
~
The baryon number density $\rho_{\rm B}^0$ inside the nucleus for the energy minimum state of the KCYN is very large, $\rho_{\rm B}^0$=(7.3$-$9.5)$\rho_0$ for H-EOS~(A) and $\rho_{\rm B}^0$=(6.2$-$8.5)$\rho_0$ for H-EOS~(B). Furthermore $\rho_{\rm B}^0$ has a maximum at around $|S|$ = 10. At the same time, the chiral angle $\theta$ for this state is saturated to the maximal value of $\pi$ (rad) irrespective of $|S|$ [the solid lines in Figs.~\ref{fig5}~(a) and (b)], corresponding to a limiting state of kaon condensation. 
Such a fully developed system of kaon condensates is naturally correlated with the highly dense state for the KCYN. 

As compared with the case of the KCYN, 
the $\rho_{\rm B}^0$ for the
KCN monotonically increases with $|S|$: $\rho_{\rm B}^0$ = 
(1.3$\rightarrow$4.7)$\rho_0$ as $|S|$ = 1$\rightarrow$20 for H-EOS~(A) and $\rho_{\rm B}^0$ = 
(1.3$\rightarrow$5.9)$\rho_0$ for $|S|$=1$\rightarrow$20 for H-EOS~(B)[the dotted lines in Figs.~\ref{fig4}~(a) and (b)]. The chiral angle $\theta$ also monotonically  increases with $|S|$ [the dotted lines in Figs.~\ref{fig5}~(a) and (b)], since the kaon condensates develop more as the baryon density gets large. 
On the other hand,
the $\rho_{\rm B}^0$ for the noncondensed hypernucleus is $\rho_{\rm B}^0$ = (1.1$-$2.1)$\rho_0$,  being much smaller than that for the KCYN for both H-EOS~(A) and (B), and the $|S|$-dependence of $\rho_{\rm B}^0$ is moderate.  

\begin{figure}[h]
\noindent\begin{minipage}[l]{0.50\textwidth}
\begin{center}
\includegraphics[height=.3\textheight]{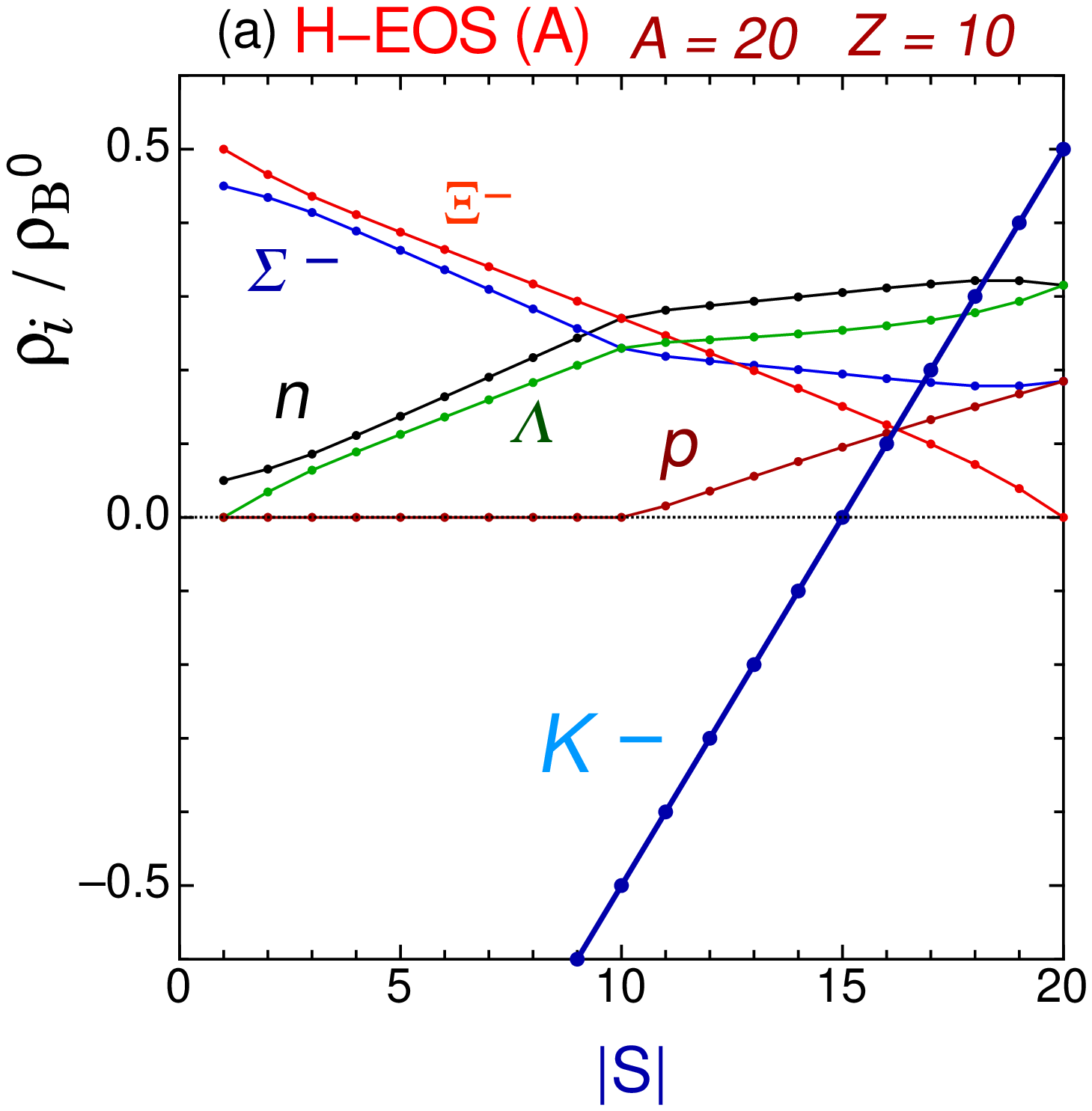}
\end{center}
\end{minipage}~
\begin{minipage}[r]{0.50\textwidth}
\begin{center}
\includegraphics[height=.3\textheight]{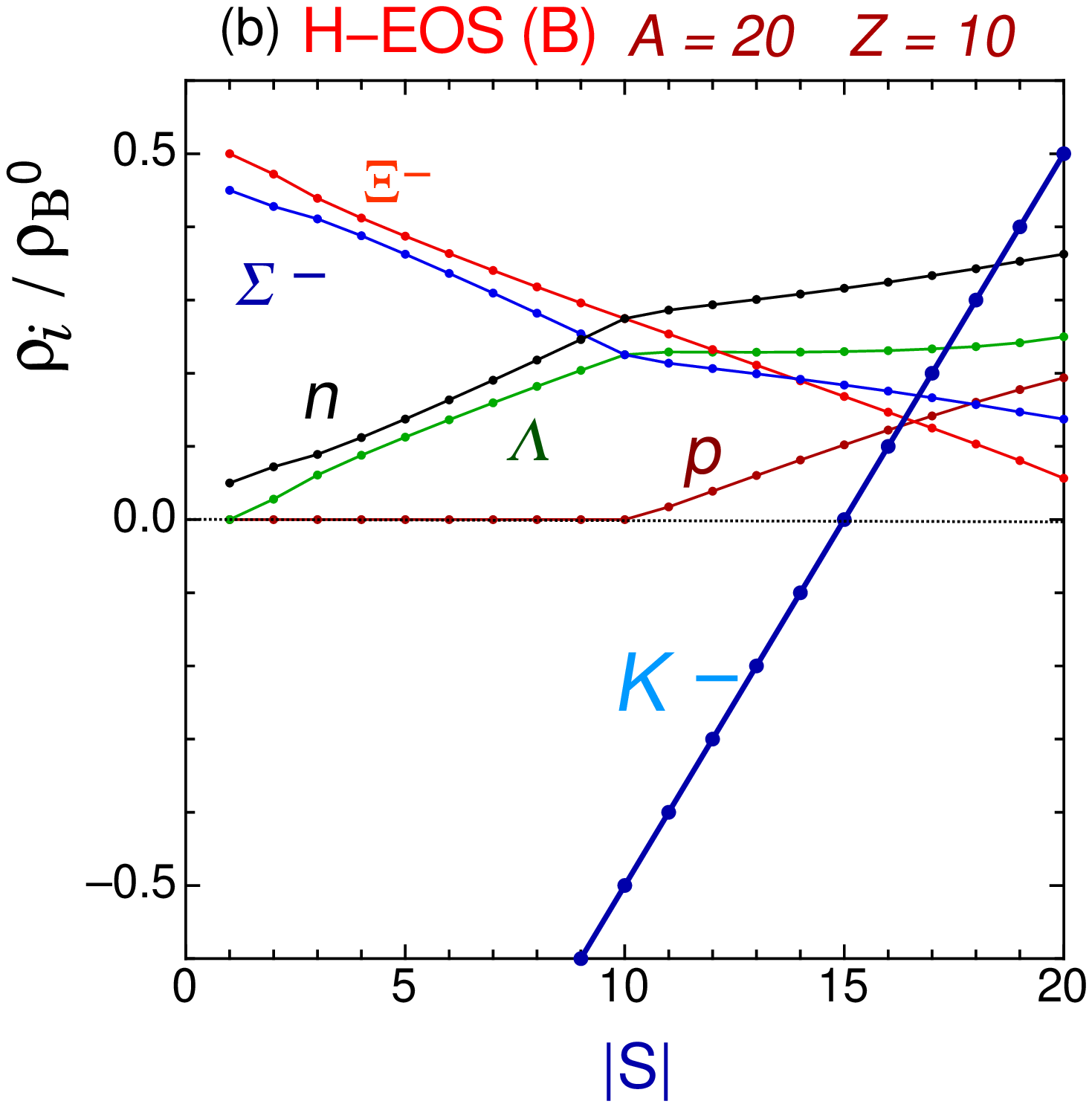}
\end{center}
\end{minipage}
\caption{(a) The particle fractions, $\rho_a/\rho_{\rm B}^0$ ($a=K^-, p, \Lambda, \Xi^-, \Sigma^-, n$), for the energy minimum state of the KCYN with $A$=20, $Z$=10 as functions of the strangeness $|S|$ for H-EOS (A) and $\Sigma_{Kn}$=305 MeV. (b) The same as in (a), but for H-EOS (B).}
\label{fig6}
\end{figure}

\subsubsection{Composition of the KCYN}
\label{subsubsec:5-2-2}

\ \ The particle fractions, $\rho_a/\rho_{\rm B}^0$ ($a=K^-, p, \Lambda, \Xi^-, \Sigma^-, n$), for the energy minimum state of the KCYN with $A$=20, $Z$=10 are shown as functions of the strangeness $|S|$ in Figs.~\ref{fig6}~(a) and (b). (a) is for H-EOS (A) and (b) is for H-EOS (B).
In the case of the KCYN, the chiral angle takes a maximal value, $\theta$ = $\pi$, for which the kaon number density $\rho_{K^-}$ reads
$\rho_{K^-}$ = $2\rho_p+\rho_n-2\rho_{\Xi^-}-\rho_{\Sigma^-}$ from Eq.~(\ref{eq:rhok}). This expression is further rewritten by the use of Eqs.~(\ref{eq:conv1})$-$(\ref{eq:conv3}) as
\begin{equation}
\rho_{K^-}/\rho_{\rm B}^0 = \left(2|S|-Z\right)/A-1 \ . \label{eq:rhokkcyn}
\end{equation}
For $|S| < (A+Z)/2$, one finds $\rho_{K^-} < 0$, which means that the strangeness and the electric charge carried by the condensed kaons become positive. Therefore for $|S|< 15$, one has effectively $K^+$ condensates. This result is kept valid irrespective of the baryon-baryon interaction models, H-EOS~(A) and (B). 

The dependence of the baryon fractions on $|S|$ for H-EOS~(A) are quantitatively similar to that for H-EOS~(B): 
For $|S|\leq 10$, the whole positive charge is carried by the kaon condensates, and the protons are absent. The positive charge is compensated by the negatively charged hyperons, $\Xi^-$ and $\Sigma^-$. The mixing ratios of both the $\Xi^-$ and $\Sigma^-$ decrease monotonically as $|S|$ increases, while the mixing ratio of the $\Lambda$ monotonically increases. For $|S| > 10$ corresponding to the deeply bound states, one has the KCYN where all the baryons ($p$, $\Lambda$, $\Xi^-$, $n$, $\Sigma^-$) are mixed.

\begin{figure}[!]
\noindent\begin{minipage}[l]{0.50\textwidth}
\begin{center}
\includegraphics[height=.3\textheight]{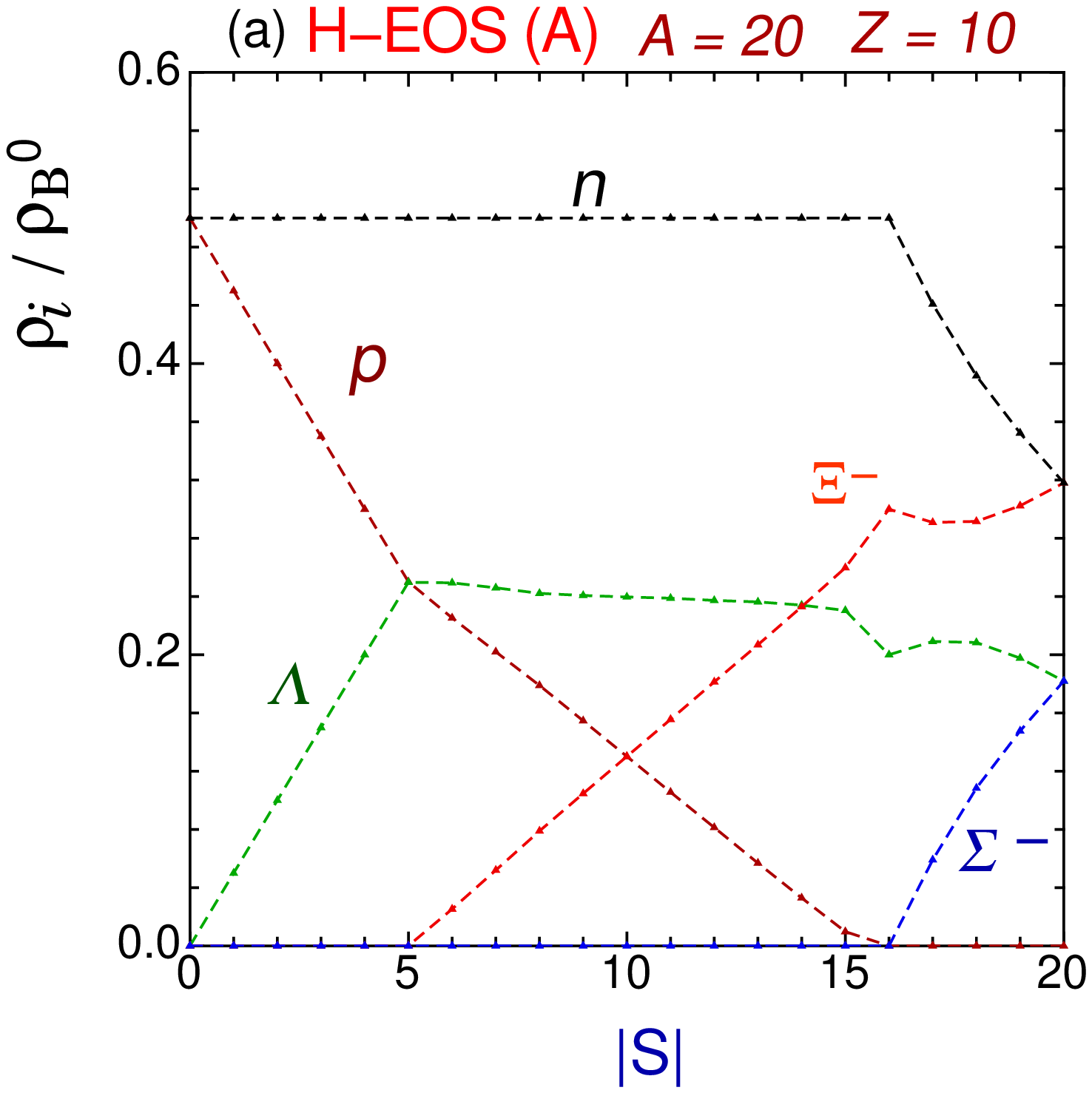}
\end{center}
\end{minipage}~
\begin{minipage}[r]{0.50\textwidth}
\begin{center}
\includegraphics[height=.3\textheight]{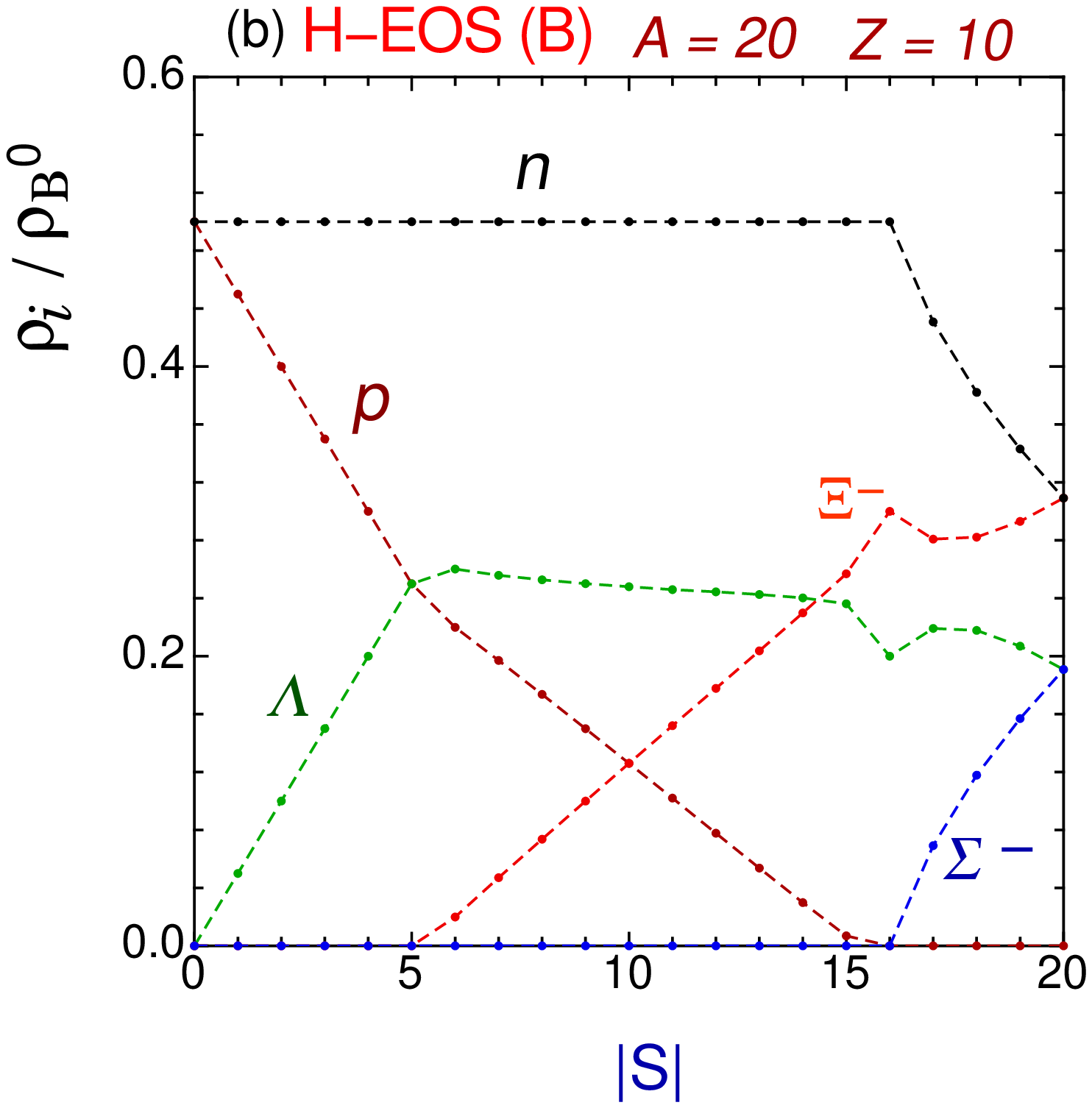}
\end{center}
\end{minipage}
\caption{(a) The particle fractions, $\rho_i/\rho_{\rm B}^0$ ($i=p, \Lambda, \Xi^-, \Sigma^-, n$), for the energy minimum state of the noncondensed hypernucleus with $A$=20, $Z$=10 as functions of the strangeness $|S|$ for H-EOS (A) and $\Sigma_{Kn}$=305 MeV. (b) The same as in (a), but for H-EOS (B).}
\label{fig7}
\end{figure}

In Figs.~\ref{fig7}~(a) and (b), we show the particle fractions for the energy minimum state of the noncondensed hypernucleus as functions of the strangeness $|S|$.
The increase in $|S|$ means that both the negative strangeness and negative charge increase. Therefore, although the mixing ratio of the proton is sizable for small values of $|S|$, it decreases monotonically as $|S|$ increases. Furthermore the $\Xi^-$ appears to be mixed at $|S|$ = 5, above which the mixing ratio of the $\Xi^-$ increases monotonically with $|S|$, while the mixing ratio of the $\Lambda$ is saturated. The $\Sigma^-$ appears to be mixed only for the large $|S|$ ($\geq 16$). 
These points are qualitatively different from the case of the KCYN, where the negative strangeness and electric  charge are carried by the kaon condensates equally as well as baryons. Hence kaon condensates lead to a drastic change of the particle composition of the nucleus. 
In the case of the noncondensed hypernucleus, the nucleus composed of the nucleons, $\Lambda$ and $\Xi^-$ is energetically stable when the strangeness fraction $|S|/A$ is large. This picture is compatible with those of the other works on the strange hadronic matter or multistrange hypernucleus within different frameworks\cite{sgs92,wssz99,sl99,shsg02}. 

\begin{figure}[h]
\noindent\begin{minipage}[l]{0.50\textwidth}
\begin{center}
\includegraphics[height=.3\textheight]{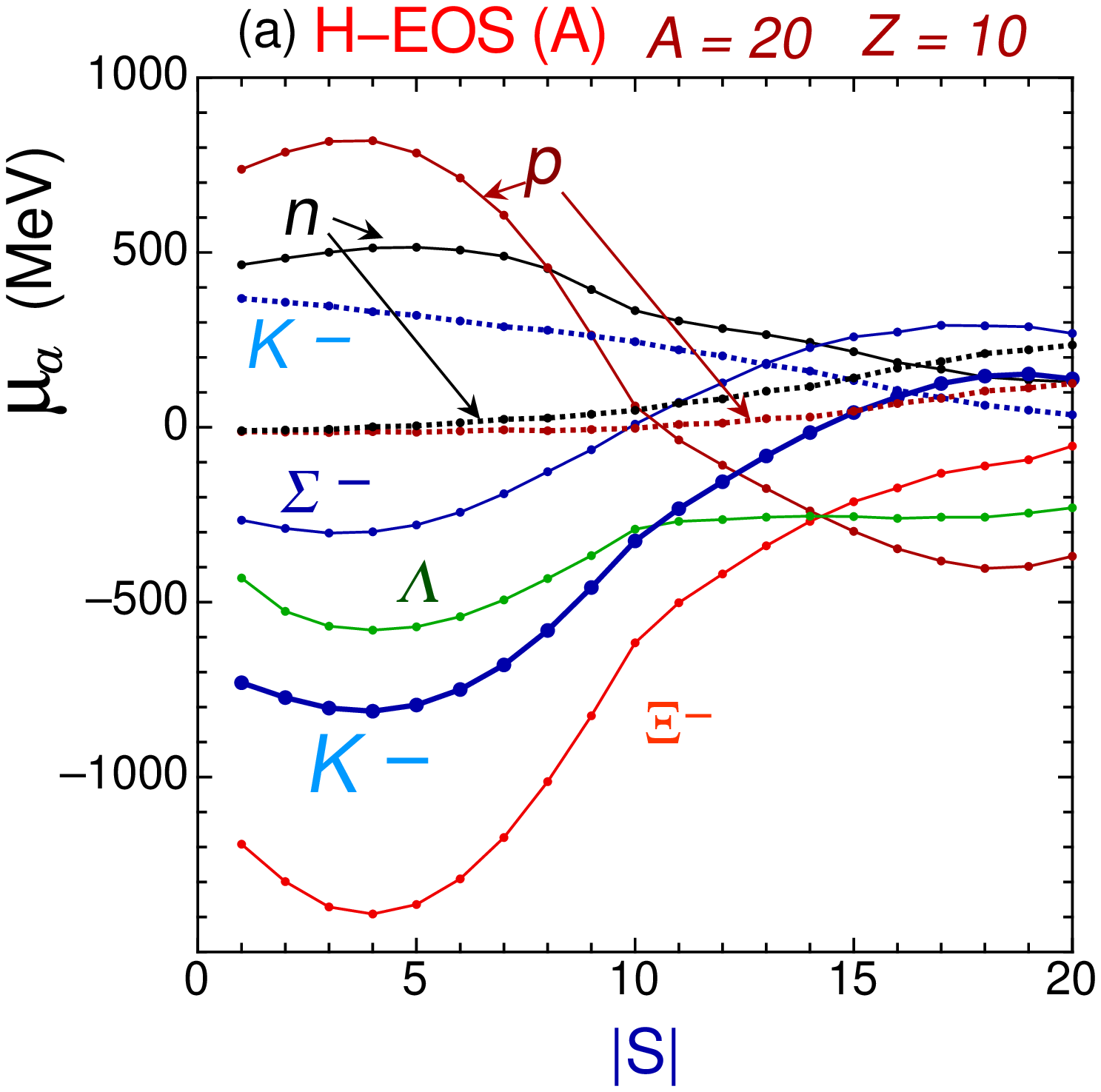}
\end{center}
\end{minipage}~
\begin{minipage}[r]{0.50\textwidth}
\begin{center}
\includegraphics[height=.3\textheight]{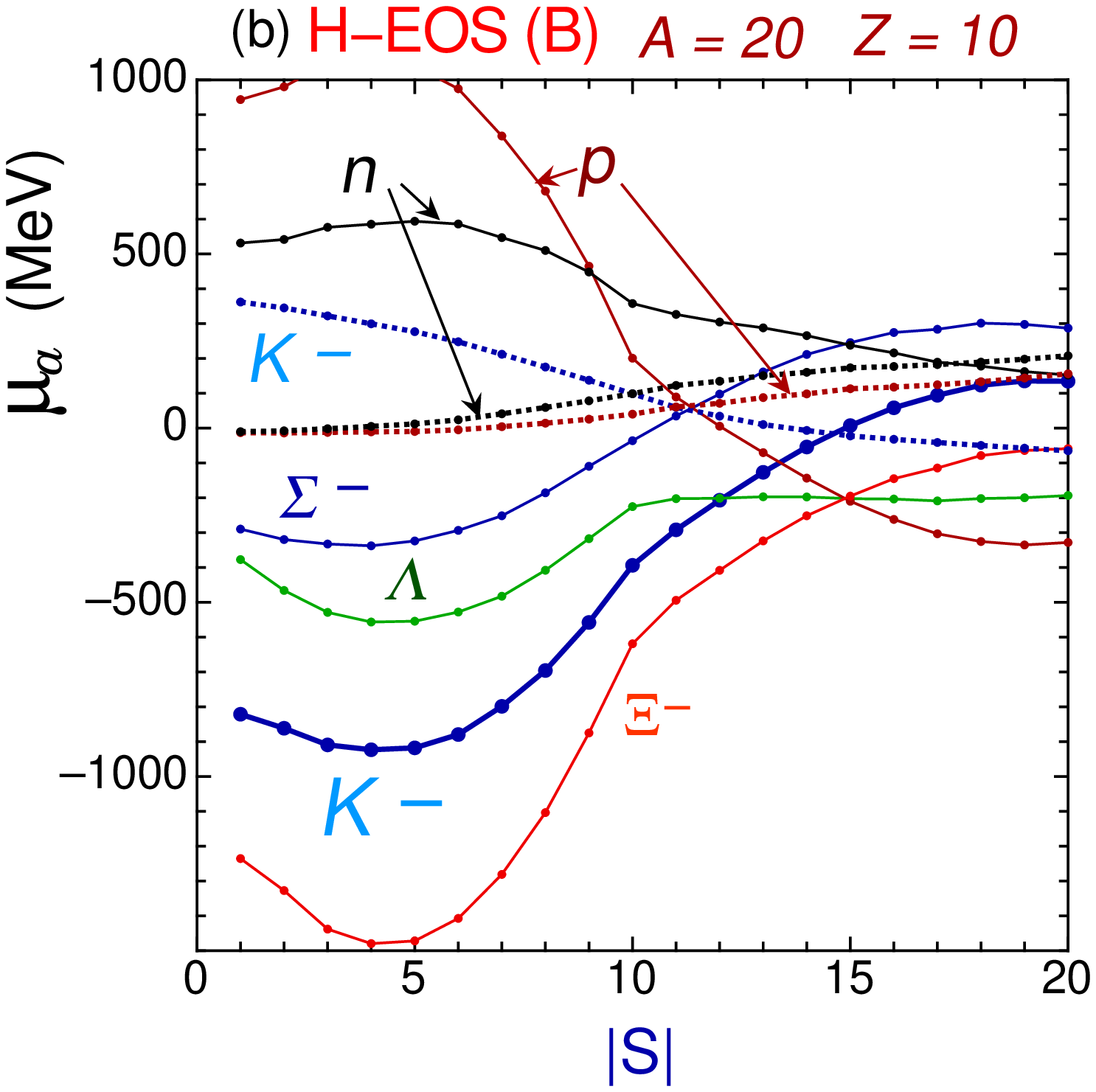}
\end{center}
\end{minipage}
\caption{(a) The chemical potentials of the classical kaon and baryons, $\mu_a$ ($a$ = $K^-$, $p$, $\Lambda$, $\Xi^-$, $n$, $\Sigma^-$), at the energy minimum state of the KCYN with $A$=20, $Z$=10 are shown as functions of the strangeness $|S|$ for H-EOS (A) and $\Sigma_{Kn}$=305 MeV by the solid lines. For comparison, those of the classical kaon, proton, and neutron at the energy minimum state of the KCN (without mixing of hyperons) are shown by the dotted lines. (b) The same as in (a), but for H-EOS (B).}
\label{fig8}
\end{figure}

The particle fractions are determined together with the chemical equilibrium conditions for the strong processes, Eqs.~(\ref{eq:sequil1})$-$(\ref{eq:sequil3}), where the chemical potentials for the classical kaon and baryons are responsible for the equilibrium conditions. In Figs.~\ref{fig8}~(a) and (b), the chemical potentials of the classical kaon and baryons, $\mu_a$ ($a$ = $K^-$, $p$, $\Lambda$, $\Xi^-$, $n$, $\Sigma^-$), at the energy minimum state of the KCYN with $A$=20, $Z$=10 are shown as functions of the strangeness $|S|$ by the solid lines. For comparison, the chemical potentials of the classical kaon, proton, and neutron at the energy minimum state of the KCN (without mixing of hyperons) are shown by the dotted lines. (a) is for H-EOS (A), and (b) is for H-EOS~(B). 
For $|S|\leq 10$, the proton-mixing ratio for the KCYN vanishes, where the Fermi momentum $p_F(p)$ in the proton chemical potential $\mu_p$ [(\ref{eq:mup})] is put to be zero. 
Also, the proton and neutron chemical potentials for the KCN (the dotted lines) are related by the relation, $\mu_n-\mu_p=\mu_Q$ with $\mu_Q$ being the electric charge chemical potential.

In the case of the KCYN (the solid lines), the kaon chemical potential $\mu_{K^-}$ increases monotonically with increase in $|S|$.
It is negative for $|S|\leq 14$, while it turns to be positive for $|S|\geq 15$. This feature can be seen from the relation $\mu_{K^-}=\mu_{\Xi^-}-\mu_\Lambda$ [(\ref{eq:sequil3})] together with Fig.~\ref{fig8}. 
On the other hand, in the case of the KCN (the dotted lines), the $\mu_{K^-}$ decreases monotonically with increase in $|S|$.
Since the kaon condensates without mixing of hyperons develop with increase in $|S|$, as seen in Sec.~\ref{subsubsec:5-2-1} (the dotted lines in Figs.~\ref{fig4} and \ref{fig5}), 
the development of kaon condensates is correlated with the decrease of the kaon chemical potential for the KCN. This feature can also be seen in the case of kaon condensation realized in neutron-star matter, where the system is in chemical equilibrium for weak interactions.  Thus mixing of hyperons also induces qualitative change of the $|S|$-dependence of the kaon chemical potentials.  

\begin{figure}[h]
\noindent\begin{minipage}[l]{0.50\textwidth}
\begin{center}
\includegraphics[height=.3\textheight]{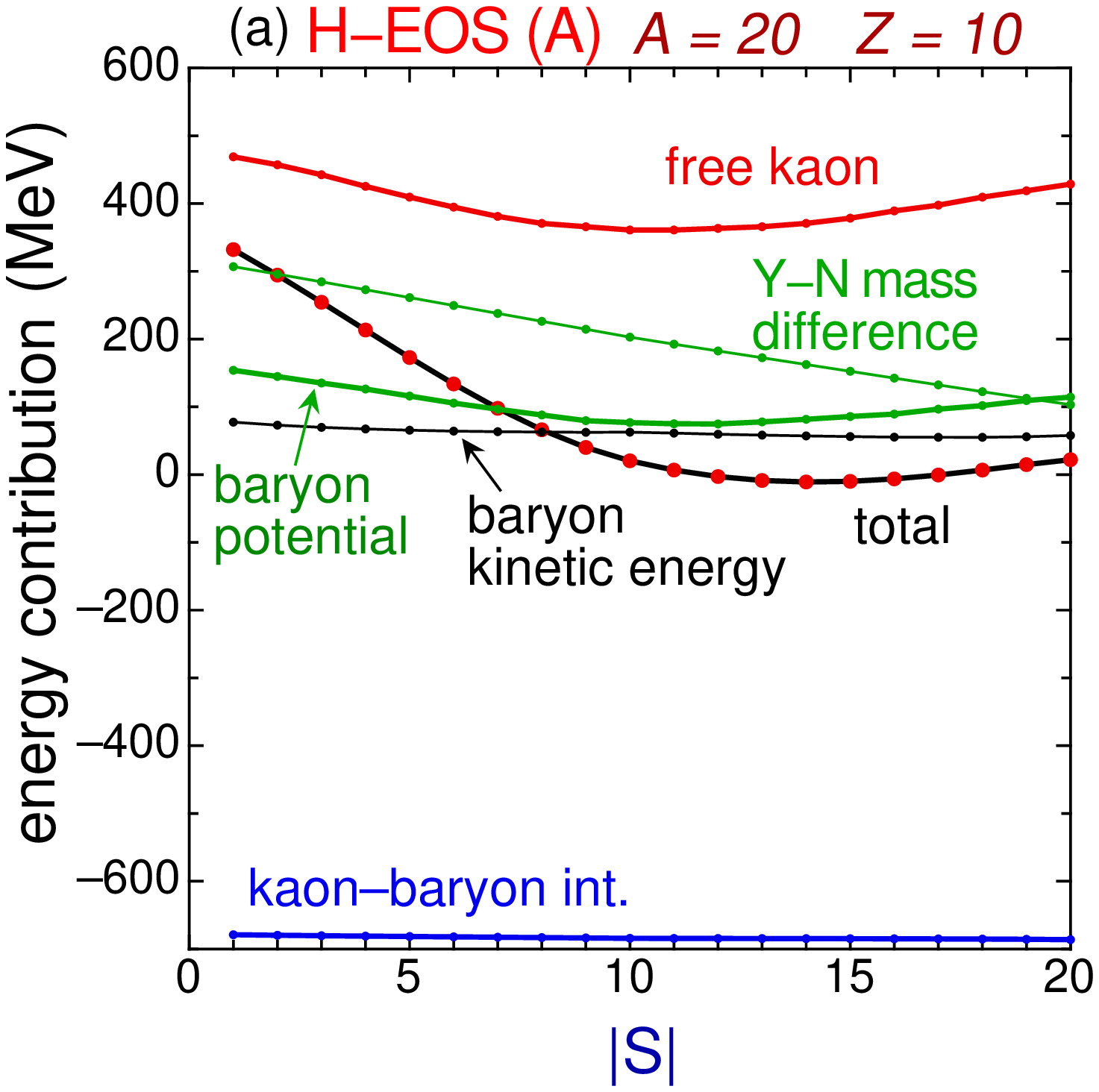}
\end{center}
\end{minipage}~
\begin{minipage}[r]{0.50\textwidth}
\begin{center}
\includegraphics[height=.3\textheight]{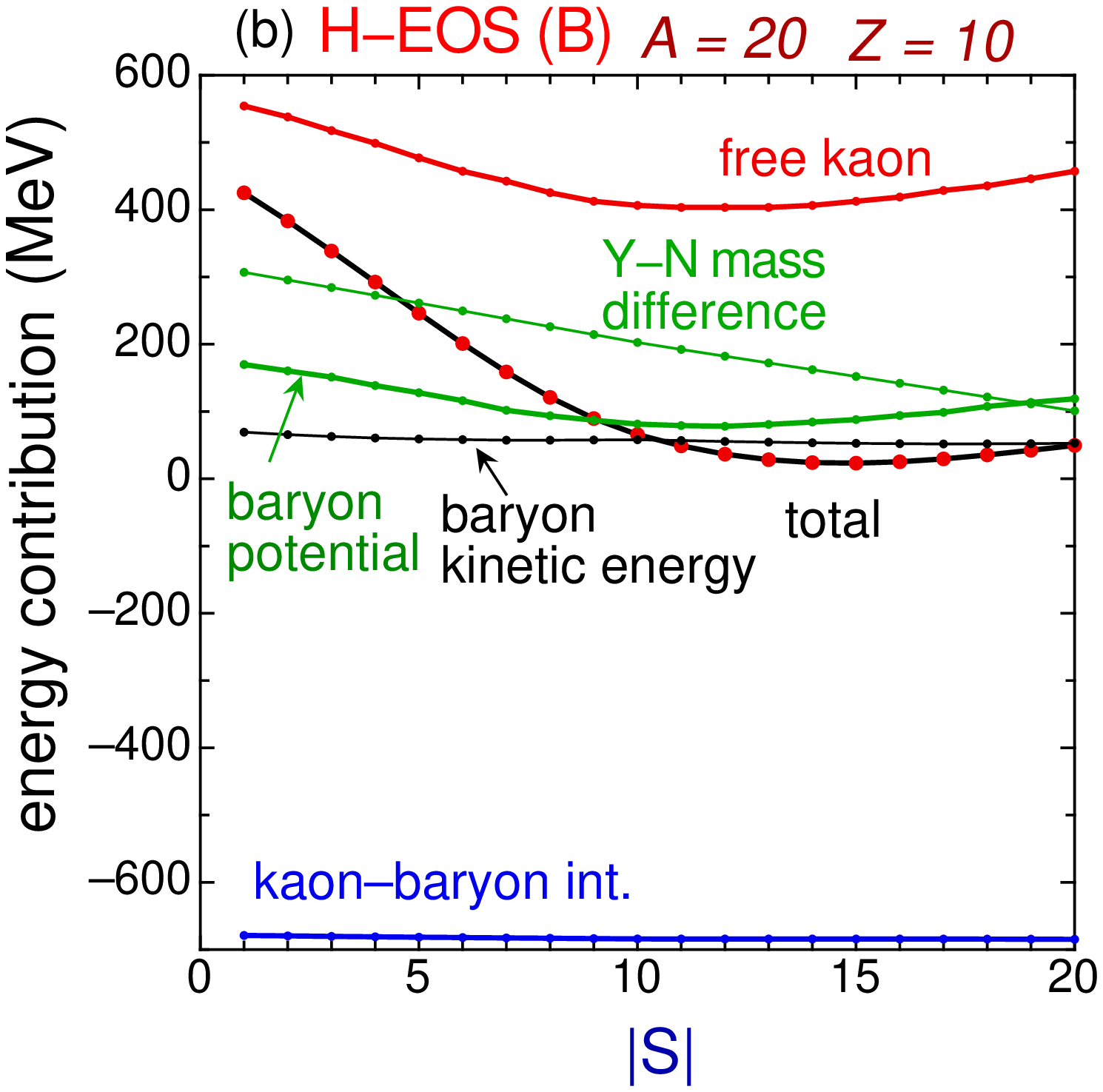}
\end{center}
\end{minipage}
\caption{(a) The contributions to the total energy per baryon for the energy minimum state of the KCYN with $A$=20, $Z$=10 as functions of $|S|$ for H-EOS (A) and $\Sigma_{Kn}$=305 MeV. (b) The same as in (a), but for H-EOS (B).}
\label{fig9}
\end{figure}

\begin{figure}[h]
\noindent\begin{minipage}[l]{0.50\textwidth}
\begin{center}
\includegraphics[height=.3\textheight]{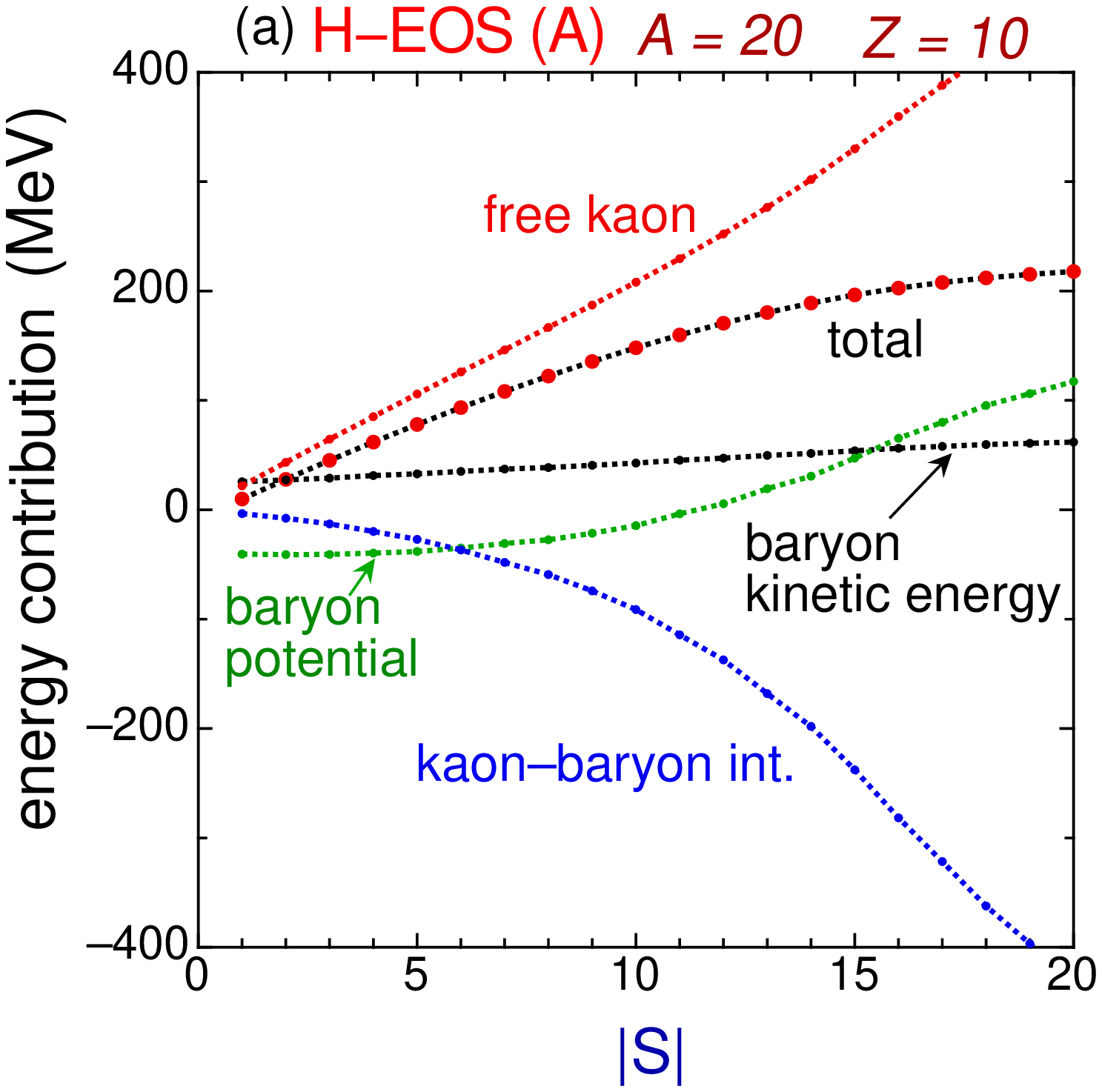}
\end{center}
\end{minipage}~
\begin{minipage}[r]{0.50\textwidth}
\begin{center}
\includegraphics[height=.3\textheight]{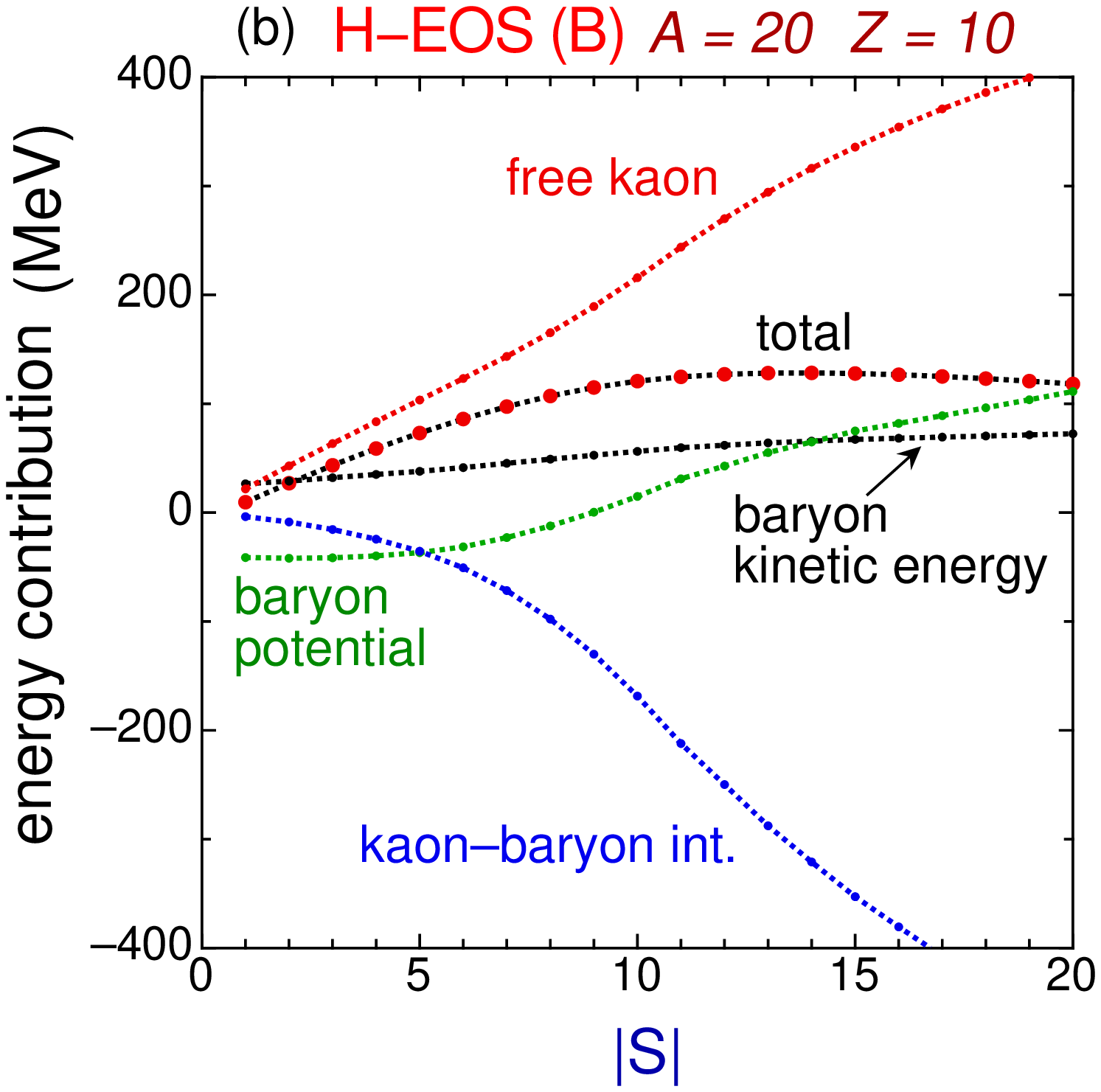}
\end{center}
\end{minipage}
\caption{(a) The contributions to the total energy per baryon for the energy minimum state of the KCN (the kaon-condensed nucleus without hyperon-mixing) with $A$=20, $Z$=10 as functions of $|S|$ for H-EOS (A) and $\Sigma_{Kn}$=305 MeV. (b) The same as in (a), but for H-EOS (B).}
\label{fig10}
\end{figure}

\subsection{Dependence of the energy contributions on $|S|$}
\label{subsec:5-3}

\ \ Here we clarify how the total energy per baryon for the energy minimum state of the KCYN depends on the strangeness number $|S|$ by analyzing each contribution to the total energy. We also compare the results for the KCYN with those for the KCN. In Figs.~\ref{fig9}~(a) and (b), 
we show the contributions to the total energy per baryon for the energy minimum state of the KCYN with $A$=20, $Z$=10 as functions of $|S|$. (a) is for H-EOS (A), and (b) is for H-EOS~(B). Those for the KCN (without mixing of hyperons) are depicted in Figs.~\ref{fig10}~(a) and (b). In all the figures, the surface energy and Coulomb energy terms are not depicted, because they are small in energy in comparison with the other terms. 
~
In the case of the KCYN, the amplitude of the kaon condensates is saturated with $\theta$ = $\pi$, and the kaon-baryon interaction term, ${\cal E}_{K-B{\rm int}}/\rho_{\rm B}^0$ [Eq.~(\ref{eq:evolkb})], is almost independent of $|S|$ and is largely negative. Also the baryon kinetic energy term, ${\cal E}_{\rm kin}/\rho_{\rm B}^0$, little depends on $|S|$. The free kaon part comes from only the kaon mass term, ${\cal E}_{{\rm free}K}/\rho_{\rm B}^0$ = $2f^2m_K^2/\rho_{\rm B}^0$ being proportional to $1/\rho_{\rm B}^0$, and has a minimum at $|S|\sim 10$,  reflecting the $|S|$-dependence of $\rho_{\rm B}^0$ (Fig.~\ref{fig4}). The hyperon-nucleon mass difference term, ${\cal E}_{Y-N{\rm mass}}/\rho_{\rm B}^0$, decreases with increase in $|S|$, due to the fact that the fractions of the heavy hyperons, $\Xi^-$ and $\Sigma^-$, decrease as $|S|$ increases (Fig.~\ref{fig6}). The dependence of the total energy per baryon on $|S|$ is mainly determined by the sum of $({\cal E}_{{\rm free}K}+{\cal E}_{Y-N{\rm mass}}+{\cal E}_{\rm pot})/\rho_{\rm B}^0$, and the total energy per baryon has a minimum at $|S|$ slightly shifted to a larger value from $|S|\sim 10$, due to addition of the 
hyperon-nucleon mass difference term which decreases with increase in $|S|$. 

In the case of the KCN (Fig.~\ref{fig10}), 
the dependence of the total energy per baryon is mainly determined by the sum of the free kaon energy, ${\cal E}_{{\rm free}K}/\rho_{\rm B}^0$, baryon potential energy, ${\cal E}_{\rm pot}/\rho_{\rm B}^0$, and the kaon-nucleon interaction term, ${\cal E}_{K-B{\rm int}}/\rho_{\rm B}^0$. As we have seen in Figs.~\ref{fig4} and \ref{fig5}, the kaon condensates for the KCN develop monotonically as $|S|$ increases, which results in that both the positive energy from ${\cal E}_{{\rm free}K}/\rho_{\rm B}^0$ and the negative energy from ${\cal E}_{K-B{\rm int}}/\rho_{\rm B}^0$ become large in magnitude as $|S|$ increases. For $|S|\sim 10$, the former conceals the latter, but above the value $|S|\sim 10$, the latter attractive effect gets more remarkable. As a result, the increase in the total energy  gets moderate for $|S|\gtrsim 10$, and the total energy per baryon becomes maximal at a large $|S|$.

\section{Implications for experiments}
\label{sec:6}

\ \ Experimental searches for deeply bound kaonic nuclear states have recently been done through the reactions, $^4$He (${\rm stopped}\ K^-,  N$)\cite{i03}, in-flight $^{16}$O($K^-$, $N$) and $^{12}$C($K^-$, $N$) reactions\cite{k05}, and stopped $K^-$ reactions on nucleus to form $\bar KNN$ or $\bar K NNN$ bound states\cite{a05}.
Producing double and/or multiple kaon clusters by way of invariant mass spectroscopy has also been proposed\cite{yda04}. 
Possible detection of the $\bar K$ nuclear clusters in heavy-ion collisions in SIS and GSI fascilities have been discussed\cite{amr06}. 
In reference to these experimental attempts, we first  briefly mention implications of our results on the KCN for experiments searching for the deeply bound multi-antikaonic nuclear clusters. Considering that the most accessible number of $K^-$ being trapped in the residual fragments in heavy-ion collisions is one or two, we show the numerical results on the baryon density $\rho_{\rm B}^0$ and binding energy per baryon $E_{\rm B}/A$ for the KCN (the case without mixing of hyperons) as ($\rho_{\rm B}^0$, $E_{\rm B}/A$) =(1.3 $\rho_0$, 6.1 MeV) for $|S|$ = 1 and (1.4 $\rho_0$, 12.8 MeV) for $|S|$ = 2, in the case of H-EOS~(A), while ($\rho_{\rm B}^0$, $E_{\rm B}/A$) =(1.3 $\rho_0$, 6.4 MeV) for $|S|$ = 1 and (1.5 $\rho_0$, 13.6 MeV) for $|S|$ = 2, in the case of H-EOS~(B). [see Figs.~\ref{fig3} and \ref{fig4}.] These states correspond to weak kaon condensation, and the density $\rho_{\rm B}^0$ inside the KCN does not attain to 2 $\rho_0$. The energy per baryon of these states lies between the lines (i) and (ii) in Fig.~\ref{fig3}, so that these states may strongly decay into mesonic or nonmesonic final states. 

Highly dense and deeply bound kaonic nucleus (KCYN) that is stable against strong interaction processes may be possible to exist as an exotic state. 
In order to obtain the KCYN in heavy-ion collisions, a number of negatively charged kaons of order $O(A)$ should be trapped in a residual nucleus. Some of the antikaons should be absorbed and hyperons should be mixed in the nucleus through the processes, $K^-N\rightarrow \pi \Lambda$, $K^-N\rightarrow \pi \Sigma$, $K^-\Lambda\rightarrow \pi \Xi$, etc. When all the remaining  antikaons occupy the lowest energy states, antikaons are expected to be condensed. Since the binding energy per strangeness of the KCYN is very large and attains to be $E_{\rm B}/|S|$ = 500 MeV (450 MeV) for H-EOS~(A) [H-EOS~(B)], a large amount of energy should be released through meson ($\pi$, $K$, $\cdots$) and other particle emissions. 

It may be statistically difficult to search for the fragments containing many antikaons in the heavy-ion collisions. Another methods for detecting the KCYN are also expected. For instance, a target nucleus could be exposed to the  high-intensity $\bar K$ beam to trap the $\bar K$ keeping stopped relative to the nucleus. 

If the KCYN is formed with a given strangeness $|S|$, it may decay into the lowest energy state with $|S|$ = 14 ($|S|$ = 15) for H-EOS~(A) [H-EOS~(B)] sequentially through weak processes [see Figs.~\ref{fig3}~(a) and(b)]. 
A typical weak reaction in the kaon-condensed state is
\begin{equation}
 N+\langle K^-\rangle\rightarrow N+e^-+\bar\nu_e     \qquad  (N = p, n)
\label{eq:ku} 
\end{equation} 
with $\langle K^-\rangle$ being the classical $K^-$ field. The reaction (\ref{eq:ku}) is a part of the kaon-induced Urca process  and is relevant to a rapid cooling of kaon-condensed neutron stars\cite{bk88, t88}. 
However, the reduced matrix elements for these processes are proportional to $\sin^2\theta$, so that they vanish in the case of the KCYN with $\theta$ = $\pi$ (rad). 
Other weak reactions such as the neutron $\beta$ decay, $n\rightarrow p+e^-+\bar\nu_e$, where the classical kaon field does not directly take part in the reactions, must be considered as possible decay modes of the KCYN. 

\section{Summary and Concluding Remarks}
\label{sec:summary}

\ \ We have considered the structure and decay modes of the kaon-condensed hypernucleus (KCYN) which may be produced in the laboratory in the strangeness-conserving processes. Our study has been based on the effective chiral Lagrangian for the kaon dynamics including the kaon-baryon interactions, combined with the nonrelativistic baryon-baryon interaction model. 

A large amount of negative strangeness, 
$|S|/A\gtrsim 0.5$, is needed in order to form the KCYN. 
In the KCYN, the kaons are maximally condensed with mixing of hyperons. The baryon density inside the nucleus is (6$ - $10)$\rho_0$ and the binding energy per strangeness, $E_{\rm B}/|S|$, attains to be $\sim$500 MeV depending on the strangeness fraction $|S|/A$ and the baryon-baryon interaction models. The incompressibility is larger than that of the normal nucleus by an order of magnitude.
It has been shown that both kaon condensates and hyperon-mixing effect cooperatively works to form such a highly dense and deeply bound object. In the case of the KCYN, the electric charge is given by $Z-|S|$, namely the positive charge of the protons is cancelled with the negative charge of the embedded $K^-$ condensates. Therefore, for a large number of strangeness such that 
 $|S| > Z$, one has a negatively charged nucleus, which may open up another way for experimental achievement of kaonic nuclear bound states.
The KCYN is stable against strong interaction processes, having a long lifetime, and may decay only through weak interaction processes.

In this paper, hadrons have been described as structureless particles. However, at high densities beyond 1 fm$^{-3}$ where the energy minimum state of the KCYN attains, the hadrons are expected to be overlapping with each other, so that the quark substructure effects might become important.\footnote{As an example, the quark-meson coupling model has been applied to discuss kaon condensation in neutron stars\cite{mpp05,rhhk07,tstw98}.}
Specifically, due to the Pauli blocking effects between quarks in each  overlapping hadrons, quarks are likely to be delocalized. Therefore, towards a more realistic consideration, one should take into account the quark substructure effects which may lead to delocalization of quarks keeping away from the repulsive interactions between baryons at high densities. 

The KCYN is characterized as a highly dense object for a large fraction of negative strangeness, $|S|/A$. This result implies a close connection with a strangelet as a dense self-bound object where asymptotically free $u$, $d$, $s$ quarks are almost equally distributed\cite{i70,ck79,w84,fj84}. The KCYN may be considered as a pathway to strange quark matter. It should be clarified whether there is a continuity between a hadronic picture for the KCYN and a quark picture for the strangelet. In a quark picture, a variety of deconfined quark phases including color superconductivity have been elaborated\cite{nhhk04}. In particular, kaonic modes ($qq\bar q \bar q$) are condensed in the color-flavor locked (CFL) phase\cite{bs02,kr02,b05,f05}. One of the issue is that condensation of such kaonic modes in the CFL phase is mediated between the hadronic and quark phases\cite{bb03}. 
For the experimental detection of the KCYN, one can expect to have common information from the experimental searches for strangelets\cite{a07}. As another viewpoint of quark substructure effects, the crossover between Bose-Einstein condensate (BEC) and BCS superconductivity in quark matter has been discussed\cite{na05}. It is instructive to investigate the patterns of  kaon condensation at high densities in relation to the BEC-BCS crossover. 

In this paper, both kinematics and interactions associated with baryons are treated nonrelativistically. The nonrelativistic EOS violates a causality at high densities. Indeed, in Ref.~\cite{m07}, 
we constructed the EOS of the kaon-condensed hyperonic matter realized in neutron stars based on the kaon-baryon and baryon-baryon interactions within the same framework as in this paper. In that case, the speed of sound exceeds the speed of light above a certain baryon density larger than that corresponding to a local energy minimum (a density isomer state) [see Fig.~12 in Ref.~\cite{m07}]. 
For more quantitative consideration, one needs a relativistic framework. Specifically, the $s$-wave kaon-baryon scalar attraction, which is proportional to the scalar densities for baryons in the relativistic framework, is suppressed at high densities due to saturation of the scalar densities\cite{fmmt96}. This effect is expected to make the EOS stiffer at high densities. 

We have assumed the uniform distribution of kaon condensates and baryons inside the nucleus.
This assumption is expected to be valid for a sufficiently large baryon number $A$. In this case, the kaon condensates may well be considered as a uniform condensation in infinite matter which may be realized in the neutron-star inner core. In accordance with this assumption within our framework, the electric charge distribution is also assumed to be uniform inside the nucleus. It is pointed out that self-consistent consideration of the finite size effects stemming from the surface tension, Coulomb interaction, and charge screening should render a non-uniform structure\cite{mtv06}. 
One should consider the validity of the assumption of the uniform distribution of particles for small values of $A$ by taking into account such finite size effects. Related work is under way with a relativistic mean-field theory\cite{mmt07}. 

Our results imply the considerably soft EOS at some density intervals as a result of coexistence of kaon condensates and hyperons\cite{m02,m07}. The underlying EOS should be consistent with observations of neutron star mass. In general, mixing of hyperons in neutron-star matter leads to a very soft nuclear EOS. The EOS based on the realistic nucleon-nucleon and hyperon-nucleon interactions with three-body forces for nucleons becomes too soft to obtain the observed canonical mass 1.44 $M_\odot$ of neutron stars with $M_\odot$ being the solar mass\cite{bbs98}.\footnote{ In this context, a phase transition to quark matter has been invoked as a solution to this problem\cite{bbs03}. } In our framework, many-body repulsive interactions between baryons (not only nucleon-nucleon but also hyperon-nucleon and hyperon-hyperon interactions) become dominant at high densities, leading to recovery of the stiffness of the EOS. 
For example, in Ref.~\cite{m02}, we applied the EOS with the H-EOS~(A) for the baryon-baryon interactions to neutron stars including a phase transition to the kaon-condensed phase in hyperonic matter and obtained a maximum mass $\sim$ 1.5 $M_\odot$.  This EOS is the same as addressed in this paper except for that both the $s$-wave and $p$-wave kaon-baryon interactions are taken into account and that the $\Xi^-$ hyperons are not included in Ref.~\cite{m02}.

However, several observational candidates suggesting large gravitational masses ($\sim$ 2 $M_\odot$ with $M_\odot$ being the solar mass) 
for neutron stars have been reported\cite{l07}. Some attempts reconciling the observations of large neutron star masses and the soft EOS associated with phase transitions have been discussed\cite{h07}. 
If it is confirmed that some neutron stars have indeed such large masses, one should consider additional repulsive effects making the EOS stiffer at very high densities. With regard to this problem, the ambiguity of the kaon-baryon and baryon-baryon interactions at high densities should be examined. It is pointed out that three-body repulsive interactions should work {\it universally} between hyperons and nucleons at high densities in  hyperonic matter in order to prevent too softening of the EOS due to mixing of hyperons\cite{nyt02}. In this context, mechanisms of the universal three-body force originating from the spin-flavor independence of the short-range repulsion has recently been considered based on the string-junction model\cite{t08}. 

\section*{Acknowledgements}
\ \ The author is grateful to A.~Gal for invitation to contribute to the special issue of the Nuclear Physics {\bf A}, ``Recent Advances in Strangeness Nuclear Physics''. He thanks T.~Tatsumi, T.~Takatsuka, T.~Kunihiro, M.~Sugawara, A.~Dote, Y.~Akaishi, and T.~Yamazaki for valuable discussions and interest in the work.
This work is supported in part by the Grant-in-Aid for Scientific Research Fund (C) of the Ministry of Education, Science, Sports, and Culture (No. 18540288) and by the funds provided by Chiba Institute of Technology.

\end{document}